\documentclass[manuscript]{acmart}

\usepackage{lib/_macros}

\AtBeginDocument{%
  \providecommand\BibTeX{{%
    \normalfont B\kern-0.5em{\scshape i\kern-0.25em b}\kern-0.8em\TeX}}}

\setcopyright{acmcopyright}
\copyrightyear{2022}
\acmYear{2022}
\acmDOI{10.1145/1122445.1122456}

\acmConference[AHs'22]{}{March 13--15, 2022}{Kashiwanoha, Japan}
\acmBooktitle{}
\acmPrice{}
\acmISBN{}

\acmSubmissionID{24}

\begin{document}

%%
%% The "title" command has an optional parameter,
%% allowing the author to define a "short title" to be used in page headers.
\title{Nightingale: AR Chat on AR Glasses}
\title{ARcall: Exploring Real-Time AR Communication for AR Glasses}
\title{ARcall: Real-Time AR Communication from Phones to Glasses}
\title{ARcall: Real-Time AR Communication using Smartphones and Smartglasses}
% \title{ARcall: Exploring Augmented-Reality-Based Real-Time Communication}

%%
%% The "author" command and its associated commands are used to define
%% the authors and their affiliations.
%% Of note is the shared affiliation of the first two authors, and the
%% "authornote" and "authornotemark" commands
%% used to denote shared contribution to the research.

\author{Hemant Bhaskar Surale}
% \authornote{}
\affiliation{%
  \institution{University of Waterloo}
  \city{Waterloo}
  \state{Ontario}
  \country{Canada}}
\affiliation{%
  \institution{Snap Inc.}
  \city{Santa Monica}
  \state{California}
  \country{USA}}
\email{hsurale@uwaterloo.ca}

\author{Yu Jiang Tham}
% \authornote{}
\affiliation{%
  \institution{Snap Inc.}
  \city{Santa Monica}
  \state{California}
  \country{USA}}
\email{yujiang@snap.com}

\author{Brian A.\ Smith}
\authornote{Co-Principal Investigators.}
\affiliation{%
    \institution{Snap Inc.}
    \city{Santa Monica}
    \state{California}
    \country{USA}}
\affiliation{%
    \institution{Columbia University}
    \city{New York}
    \state{New York}
    \country{USA}}
\email{bsmith@snap.com}

\author{Rajan Vaish}
\authornotemark[1]
\affiliation{%
    \institution{Snap Inc.}
    \city{Santa Monica}
    \state{California}
    \country{USA}}
\email{rvaish@snap.com}

% \author{Ben Trovato}
% \authornote{Both authors contributed equally to this research.}
% \email{trovato@corporation.com}
% \orcid{1234-5678-9012}
% \author{G.K.M. Tobin}
% \authornotemark[1]
% \email{webmaster@marysville-ohio.com}
% \affiliation{%
%   \institution{Institute for Clarity in Documentation}
%   \streetaddress{P.O. Box 1212}
%   \city{Dublin}
%   \state{Ohio}
%   \postcode{43017-6221}
% }

% \author{Lars Th{\o}rv{\"a}ld}
% \affiliation{%
%   \institution{The Th{\o}rv{\"a}ld Group}
%   \streetaddress{1 Th{\o}rv{\"a}ld Circle}
%   \city{Hekla}
%   \country{Iceland}}
% \email{larst@affiliation.org}

% \author{Valerie B\'eranger}
% \affiliation{%
%   \institution{Inria Paris-Rocquencourt}
%   \city{Rocquencourt}
%   \country{France}
% }

% \author{Aparna Patel}
% \affiliation{%
%  \institution{Rajiv Gandhi University}
%  \streetaddress{Rono-Hills}
%  \city{Doimukh}
%  \state{Arunachal Pradesh}
%  \country{India}}

% \author{John Smith}
% \affiliation{\institution{The Th{\o}rv{\"a}ld Group}}
% \email{jsmith@affiliation.org}

% \author{Julius P. Kumquat}
% \affiliation{\institution{The Kumquat Consortium}}
% \email{jpkumquat@consortium.net}

%%
%% By default, the full list of authors will be used in the page
%% headers. Often, this list is too long, and will overlap
%% other information printed in the page headers. This command allows
%% the author to define a more concise list
%% of authors' names for this purpose.
\renewcommand{\shortauthors}{Surale, et al.}

%%
%% The abstract is a short summary of the work to be presented in the
%% article.

\begin{abstract}

Augmented Reality (AR) smartglasses are increasingly regarded as the next generation personal computing platform. However, there is a lack of understanding about how to design communication systems using them. We present ARcall, a novel Augmented Reality-based real-time communication system that enables an immersive, delightful, and privacy-preserving experience between a smartphone user and a smartglasses wearer. ARcall allows a remote friend (\emph{Friend}) to send and project AR content to a smartglasses wearer (\emph{Wearer}). The ARcall system was designed with the practical limits of existing AR glasses in mind, including shorter battery life and a reduced field of view. We conduct a qualitative evaluation of the three main components of ARcall: \emph{Drop-In}, \emph{ARaction}, and \emph{Micro-Chat}.
%Drop-In lets the Friend see what the Wearer is seeing before starting an AR call, to probe the wearer\rq s context. ARaction lets the Friend project AR content in the Wearer’s environment. Micro-Chat lets the Friend experience the Wearer\rq s reactions through a time-bounded voice call.
Our results provide novel insights for building future AR-based communication methods, including, the importance of context priming, user control over AR content placement, and the feeling of co-presence while conversing.

% The AR contents, placed in the real-world, can  augment the Wearer's reality. 
% We conduct a preliminary evaluation of our system to validate the three components of ARcall. Our results suggest that Friends felt that they were augmenting Wearers\rq reality, while Wearers felt that the AR content from their Friends was immersive, relevant, and personalized.

\end{abstract}

%%
%% The code below is generated by the tool at http://dl.acm.org/ccs.cfm.
%% Please copy and paste the code instead of the example below.
%%

\begin{CCSXML}
<ccs2012>
<concept>
<concept_id>10003120.10003138.10003140</concept_id>
<concept_desc>Human-centered computing~Ubiquitous and mobile computing systems and tools</concept_desc>
<concept_significance>500</concept_significance>
</concept>
</ccs2012>
\end{CCSXML}

\ccsdesc[500]{Human-centered computing~Ubiquitous and mobile computing systems and tools}

%\ccsdesc[500]{Human-centered computing~HCI theory, concepts and models}
% \ccsdesc[500]{Human-centered computing~Empirical studies in HCI}

%%
%% Keywords. The author(s) should pick words that accurately describe
%% the work being presented. Separate the keywords with commas.
\keywords{augmented reality, immersive communication, calling}

\begin{teaserfigure}
  \includegraphics[width=\textwidth]{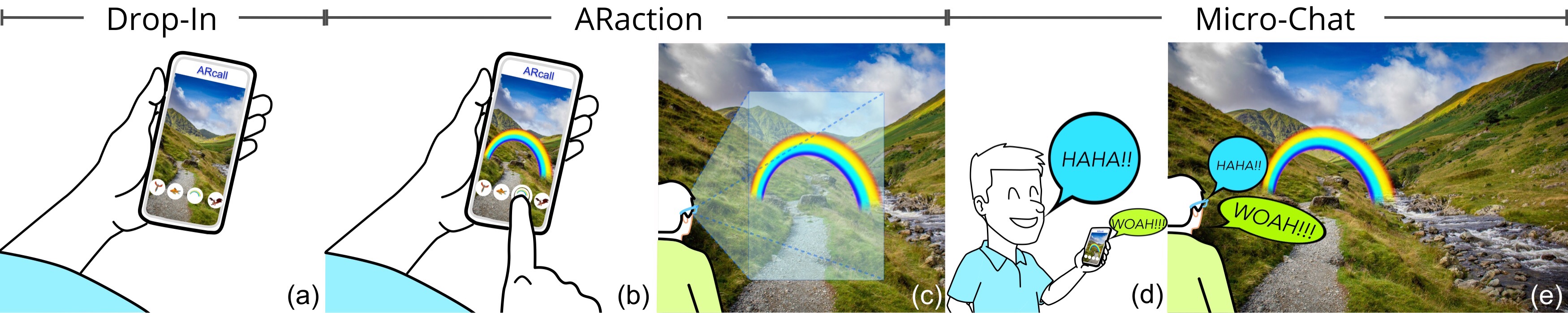}
  \caption{ARcall's main components, which occur in real-time, including Drop-In, ARaction and Micro-Chat: (a) Drop-In allows the Friend to see the Wearer\rq s view on a smartphone; (b-c) ARaction allows the Friend interact with the Wearer by sending AR content; (b) The Friend selects an AR rainbow to send; (c) The Wearer sees the rainbow augmenting their reality; (d-e) Micro-Chat allows the Friend and the Wearer to talk to each other; (d) The Friend says "Hahaha!!" after sending the rainbow; (e) The Wearer reacts to the rainbow by saying, "Whoa!!" as the Friend remains dropped in.}
 \Description{}
%   \Description{Enjoying the baseball game from the third-base
%   seats. Ichiro Suzuki preparing to bat.}
  \label{fig:teaser}
\end{teaserfigure}

%%
%% This command processes the author and affiliation and title
%% information and builds the first part of the formatted document.
\maketitle

\section{Introduction}

As AR-based smartglasses are evolving, they are unlocking exciting new forms of immersive communication.
Prior research has explored the utility of AR mainly as a means of annotating environments~\cite{Gauglitz2014,Jo2013,Kim2019,Nassani2016}, communicating with avatars~\cite{piumsomboon2018mini}, relaying remote instructions and guiding \cite{thoravi2019loki, henderson2011augmented}, visualizing biosignals~\cite{Piumsomboon2017}, and rendering visual alterations \cite{rixen2021exploring}. However, there is a lack of research in understanding how AR serves in the context of informal communication using modern AR smartglasses. \cite{SnapSpectacles}. Specifically, Pfeil \ea~ \cite{pfeil2021bridging} 
highlighted the need for telepresence systems to consider both streamers' and viewers' experiences when communicating with each other via devices such as smartglasses 
and via device pairings such as smartglasses and smartphones.
Given the recent surge in AR smartglasses in the consumer market \cite{nreal, focalssetup, GoogleGlass, VuzixBlade, Rokid}, it is evident that everyday communication practices will extend to AR smartglasses  \cite{pfeil2021bridging}. Therefore, we explore the design of ARcall, an Augmented Reality (AR) based communication system, in the context of informal communication between a smartphone user and a smartglasses wearer.

We introduce ARcall, a native AR-based real-time communication system designed to support the one-to-one paradigm \cite{pfeil2021bridging}. Our system consists of two software applications: a smartglasses application for the smartglass user (the \quotes{Wearer}) and a smartphone application for the smart phone user (the \quotes{Friend}). ARcall enables sending and projecting AR content to the Wearer \textit{natively} in AR format. For instance, a Friend can send AR content like making it snow or dropping a rainbow into the Wearer's environment (Figure \ref{fig:teaser} (b--c)). When designing ARcall, we considered the benefits of wearable and ready-to-project AR smartglasses, in addition to socio-technical constraints such as privacy, interruptions, and mixed-device configuration necessary for interactions between smartphones and smartglasses \nolinebreak\cite{ackerman2000intellectual}.

ARcall consists of the following main design components:
% \vspace{-3mm}
\begin{itemize}
 \item \textit{Drop-In}: Drop-In allows the Friend to see the Wearer's current context (i.e., where they are and what they are doing) by \quotes{dropping in} and viewing a video feed of the Wearer’s point of view (see Figure \ref{fig:teaser}~(a)). Drop-In is an invite-only feature in which the Friend is granted access to time-limited \textit{Drop-In sessions} only when the Wearer goes online and invites them. Note that the Friend can decide to Drop-In at their convenience during the session.

 \item \textit{ARaction}: ARaction enables interaction with the Wearer using AR. While dropping in on the Wearer from the smartphone, the Friend can browse, select, and send AR content to the Wearer (see Figure \ref{fig:teaser}~(b--c)). The AR content is then displayed on the Wearer’s glasses, augmenting their reality in real time.
 
 \item \textit{Micro-Chat}: A Micro-Chat is a time-bounded voice call between the Wearer and Friend. Micro-Chats enable the Friend to add meaning to the AR content they send and experience the Wearer’s reaction in real time (see Figure \ref{fig:teaser}~(d--e)). While talking, the Friend can continue to send AR content and see the Wearer's point of view (with AR content they've just sent superimposed onto it) in real time. The Micro-Chat is capped at one minute and happens as part of the Drop-In session. It runs for as long as Drop-In session lasts, which preserves the battery life of the AR glasses. The Wearer can extend the Drop-In session in 30-second increments by tapping a button on the temple of the smartglasses during the session.

\end{itemize}

% \vspace{-3mm}
We conducted a qualitative evaluation of ARcall with 14 participants to probe overall experience, initial perceptions of the system, and feelings of connectedness and understand the limitations of native AR communication. We learned that both Friends and Wearers found ARcall to be fun and immersive and that ARaction added an element of surprise to their interactions. Friends felt that they were augmenting Wearers' reality, and Wearers felt that the AR content that their Friends sent was relevant and personalized to their environment. Wearers felt connected to their Friend by giving the Friend the ability to project AR content to their display. ARcall's Drop-In component helps Friends learn about the Wearer's context (helping the Friends select and send meaningful, personalized AR content), seeing the Wearer's point of view also increased the users' feeling of togetherness. Finally, ARcall's Micro-Chat component created intimacy and added meaning to the interaction. As soon as Wearers heard their Friends talk, they felt they were sharing a moment together. Friends highly valued the ability to experience the Wearer's reaction in real-time.

% \smallbreak
\noindent
The main contributions of this paper are as follows:

% \vspace{-3mm}
\begin{itemize}
  \item  Understanding user behaviour and the affordances created by our augmented-reality-based real-time communication method. Our design approach takes into account the practical limitations of modern smartglasses.
  
  \item Our qualitative evaluation of the key components of our proposed system provides novel insights for building future AR-based communication methods. For instance, our evaluation revealed the importance of context priming using Drop-In, engagement using ARaction, and creating meaningful conversations using Micro-Chat.
\end{itemize}

\section{Background and Related Work}

We examined several threads of communication-related research from the perspective of designing communication systems for AR smartglasses: AR view annotation, pre-call context assessment, and time-bounded calling.

\subsection{AR View Annotation} 

% Over the years, video calls have improved to support multi-modal interactions. For instance, text-based chats overlaid on a video \cite{tang2016meerkat, weisz2007watching}. However, text chats have limited expressiveness, especially, when they refer to a specific moment or objects in the video stream \cite{yang2020snapstream}, and the visual interaction like annotating a snapshot of the stream and sharing the annotated snapshots make the interaction lengthier \cite{yang2020snapstream, hamilton2018collaborative, kim2018using}. These interactions involve multiple steps that break the flow and reduces the feeling of togetherness during the call.

% , johnson2015handheld, bai2020user
% fakourfar2016stabilized, kim2014improving

To make video calls more interactive, researchers have explored how AR-based annotations can assist users engaged in collaborative tasks~\cite{gauglitz2014touch, hollerer1999exploring, billinghurst2014social, oda2015virtual}. Jo \ea{}~\cite{jo2013chili} investigated a smartphone-based system that allows a user to control a remote scene and augment it with drawings during a live video call. Similarly, Ryskeldiev \ea{}~\cite{ryskeldiev2018streamspace} introduced a system capable of sharing the smartglasses wearer's point of view (POV), including photospherical imagery and the wearer's current orientation. The system helped reduce the cognitive load of peers involved in a remote collaboration task. Further, Nassani \ea{}~\cite{nassani2016augmented} found that anchoring text on live video in a spatialized way makes them more effective than displaying static text. Past research focused on creating shared experiences for collaborative tasks using either smartphones or desktop computers~\cite{kuzuoka1994gesturecam, kim2019evaluating, muller2016panovc}, but it is unclear how these results would translate to smartglasses-based communication. Users exhibit different behaviors when video calling on smartglasses than when they use smartphones. For instance, Kun \ea{}~\cite{kun2019calling} show that smartglasses wearers do not spend time looking at callers' video feeds. Similarly, in virtual reality environments, He~\ea{}~\cite{he2020collabovr} show that the communication typically goes beyond conducting face-to-face communication or collaborative tasks to projecting shared artifacts with remote audiences. In summary, most of the past research focuses on \textit{collaboration}; the extent to which these results can translate to native AR \textit{communication} is unknown. 

Furthermore, current AR-based communication practices are limited in four ways \cite{snapagelens, smoodji, Graffity}. First, the sender can send AR content, often in the form of a pre-recorded video with superimposed AR content, without seeing the Friend's current context, so the AR may not be relevant to the Friends' surroundings. Second, these experiences are less mobile-friendly since users have to hold their smartphones facing themselves. Third, they are suitable only for smartphones or desktop computers rather than smartglasses. ARcall, on the other hand, focuses on providing users with the ability to remotely augment the reality of the smartglasses wearer in real-time, as opposed to sending pre-recorded AR content or streaming a 3D model of the caller using a complex assembly to emulate face-to-face communication \cite{fuchs1994virtual, kim2012telehuman, orts2016holoportation}.

% \rajan{we need a transition to AR glasses, and mention the challenge that we are addressing/solving}\hem{addressed this.}
% \rajan{We need to highlight the fact that we cannot copy paste things from phone to glasses}\hem{addressed this.}

% Especially, extending the previous methods to facilitate native AR-based communication on smartglasses is still an open research question. Current AR-based communication are limited in three ways. At first, they mandate the user to send a pre-recorded video with AR content as a message \cite{snapagelens, smoodji}. Secondly, these experiences support for limited mobility situations. And at last, they are suitable only for devices like a smartphone or a desktop. 

% Supporting live AR augmentation during a video call on smartglasses is interesting for several reasons. Unique scenarios like letting a remote friend augment your reality, to let them surprise you, or induce emotions, or to tell you a story using AR content could greatly improve the experience of using smartglasses. We aim to fill this research gap.

\subsection{Context Assessment before Calling} 
% A typical start of an unplanned audio or a video call begins with questions like "Are you busy?", "Where are you?", "Is it a good time to talk?".
Allowing friends or family members to bring unique experiences to users' smartglasses in a satisfying way without any explicit triggers from them is a challenge. It is important for callers to know their friends' availability before calling and potentially interrupting them. Ames et al.~\cite{ames2010making}, however, show that family members often do not schedule video calls in advance and consider it socially strange or unnatural to do so. Unplanned audio and video calls typically start with questions such as "Are you busy?", "Where are you?", and "Is this a good time to talk?"; however, in the AR context, it may be too late to ask such questions once the call has been placed. Nardi \ea{}~\cite{nardi2000interaction} highlighted this fundamental asymmetry in conversation, referring to limits on the receiver's availability, especially when the person is engaged in another task or conversation.

One way to tackle such issues is by implementing a two-step calling process. In the first step, a caller gauges the context of the callee and in the next step, initiates a call. For instance, Cramer \ea~\cite{cramer2011performing} investigated the "check-in" model of location sharing, in which a user pre-defines a list of friends or publicly shares their check-in location to help others determine an appropriate time to call. A similar study by Schildt \ea{}~\cite{schildt2016communication} reported that the participants often checked the location of someone they were about to contact to determine if the person they wanted to call was available for further communication. Pfleging \ea{}~\cite{pfleging2013exploring} proposed a solution that shares a driver's live camera feed; this enables the caller to become a "virtual passenger" in the vehicle, thereby empowering the caller to gauge the current situation and act accordingly. Further, modern video chatting applications address this challenge by notifying others when the callee is available or online. For instance, apps like Houseparty app~\cite{houseparty} and Clubhouse~\cite{clubhouse} send users notifications when their friends are available for a video or audio chat. Facebook Rooms~\cite{FBRooms} adopts a similar approach. 

Procyk \ea{}~\cite{procyk2014exploring} prototyped a wearable video chat experience in which loved ones participated in a collaborative task over distance. Participants used smartphones and cameras mounted on either a hat or a normal pair of glasses. However, to date, no past work has explored the role of context assessment prior to initiating a conversation with a smartglasses wearer, an aspect of AR communication we investigate with the ARcall\rq s Drop-In feature. %Especially, when modern smartglasses could share their POV with the caller in real-time. In native AR communication, such context plays an important role for the caller. The caller can decide to send AR content to the smartglass wearer that are dependent on their immediate reality, than based on a limited location information.

% \rajan{We got to mention that because AR augments the physical world, it becomes critical to understand the context even more --- its not just about what the person is doing, but where are they --- are they indoor or outdoor etc --- so that relevant lenses can be launched}\hem{Added it.}\hem{Done.}

\subsection{Time-Bounded Calls for Mobility}
Video chat as a means of experience sharing has received significant attention ~\cite{baishya2017your, neustaedter2020mobile, greenberg2013shared}; the use of AR glasses to support video-based exchanges is a particularly attractive extension, as smart eyewear can support hands-free experience sharing with great ease. Typically, such scenarios involve the smartglasses wearer broadcasting their video and the remote user viewing the video and participating virtually. Increasing the remote user's sense of participation is key to making such interactions engaging. Procyk \ea{}~\cite{procyk2014exploring} showed that audio played a central role in creating a strong sense of presence and connection with the remote partner during a shared remote activity. Similarly, Neustaedter \ea{}~\cite{neustaedter2015sharing} examined the role of audio during a passive always-on video call in which family members simply wanted to be aware of what another member was doing without explicitly conversing during the call. Depending on the situation, an AR glasses wearer may not always want to see a video of the caller~\cite{kun2019calling}.

Despite the advantage of AR smartglasses allowing others to share the wearer's viewpoint anytime and having a screen that can project anytime, they also have a significant disadvantage in terms of experience sharing. Specifically, voice and video calls drain their batteries quickly, rendering them unusable until they are recharged. One potential solution to this problem is to support time-bound calls on AR smartglasses. Limiting call length may also help the wearer to focus, as past studies emphasize that users' cognitive resources are very limited when they are mobile~\cite{oulasvirta2005interaction}. There is a lack of significant research on time-bounded video calls using AR glasses.

\smallbreak
We attempt to provide a feasible solution to the three core problems outlined above --- enhancing interaction, enabling context assessment, and implementing time-bounded calling to limit power consumption --- in designing for a wearable form factor. With ARcall, our goal is to reinvent traditional communication practices for wearable smartglasses, Keeping in mind the challenges such as shorter call duration (due to limited battery), context assessment, and AR-based interactions, Pfeil \ea~ \cite{pfeil2021bridging} emphasized the need to develop a new set of design considerations for use with modern wearable technologies. Therefore, we will analyze a unique set of design considerations in the following section.

\begin{comment}

https://arinsider.co/2018/01/04/bringing-ar-to-messaging-a-conversation-with-snaappy/

https://medium.com/@Smoodji/smoodji-a-revolutionary-ar-app-a1f5294fd53b

https://medium.com/inborn-experience/ar-video-chat-graffity-52cdc87226a1

\end{comment}

\section{Design Considerations}

Our design considerations focus on the benefits of the mobile, wearable, and ready-to-project AR smartglasses form factor while also accounting for socio-technical limitations \cite{ackerman2000intellectual} such as privacy, interruptions, control, and so forth.

% % % % % % % % % % % % % % % % % % % % % % % % % 
% \medbreak
\noindent
%\textbf{Before the ARcall:}

\begin{design}\label{design:interrupt} \textit{Giving control to Wearers and freedom to Friends:}
An AR communication system must balance  the interests of both the Wearer and the Friend: it must give Friends (callers) the ability to initiate calls ~\cite{oulasvirta2005interaction} while preserving the Wearer's ultimate control over when they can be called~\cite{cutrell2000effects}. Through ARcall's \textit{Drop-In} component, we explore two techniques for addressing these competing interests. First, Drop-In is invite-only, meaning Friends can only drop in when the Wearer starts a session and invites them. Second, once invited, the Friend can drop in at any time to gauge the Wearer's context --- where they are and what they are doing --- \textit{before} sending AR content and speaking to the Wearer in a Micro-Chat. Once invited, the Friend can drop in as many times as they want during the ARcall session. 

%the process of dropping in allows the Friend to discover what the Wearer is doing \textit{before} interrupting them with AR content, and gauge whether it is a good time to start an ARcall.
\end{design}

\begin{design}\label{design:privacy} \textit{Respecting the Wearer\rq s privacy:} Sharing one's real-time camera feed with a remote Friend can make smartglass communication more compelling, but it also raises privacy concerns. Often, users have concerns over whether they might be bothering the person they want to reach before placing a call~\cite{judge2010sharing}. Our design for ARcall explores three techniques for mitigating privacy concerns. First, ARcall's \textit{Drop-In} component is invite-only, as described in D1, and is designed only for interactions with a single Friend. Second, invitations automatically expire after one hour. Third, ARcall gives Wearers the option of blurring their video feeds. Naively, one might be tempted to include additional privacy controls, but it is critical to balance explicit efforts to fine-tune the privacy controls against the risk of degrading the user experience \cite{ackerman2005privacy}.

\end{design}

% \medbreak
\noindent
%\textbf{During the ARcall:}

\begin{design}\label{design:lenses} \textit{Facilitating relevant and well-placed AR content:} 
Native AR communication is potentially compelling for users because it can directly augment the Wearer's reality. For example, a Friend might send AR balloons to the Wearer on their birthday, and the balloons might appear to float in front of them. Two key determinants of whether or not the AR is compelling, however, are whether it is well-suited to the Wearer's current environment (i.e., whether it is relevant to the Wearer's context \cite{louie2021opportunistic}), and whether it is well-placed within that environment \cite{ran2019sharear}. For instance, in ARcall, Drop-In allows the Friend to see that the Wearer is celebrating their birthday, and ARaction allows them to augment the Wearer's reality with a floating balloon in the same way they could in the real world.

\end{design}

\begin{design}\label{design:fov} \textit{Field-of-view accommodation:} Effective AR experiences must take into account the limited field of view afforded by the displays currently built into smartglasses. 
With this in mind, in designing ARcall, we explored the effectiveness of restricting AR content to allow 3D objects (virtual models) only rather than entire \quotes{world reskins} \cite{krawczyk2005perceptual, danieau2017attention}. In our exploration, Wearers found the AR content to be more compelling when it augmented \textit{part} of their environment rather than the \textit{entire} environment.

\end{design}

\begin{design}\label{design:power} \textit{Handling power constraints:} Power constraints are a major hurdle for current smartglass devices \cite{hashimoto2016blinded, matsuhashi2020thermal}. This is especially true for AR communication applications: prolonged and simultaneous operation of smartglass device displays, microphones, and live camera feeds (for instance, during audio or video calls) mean that it is critical to carefully manage power consumption.
To constrain power consumption while giving users flexibility, Drop-In sessions are kept short: the Wearer can configure them to last between 30 seconds and one minute. However, if desired, the Wearer can extend active Drop-In sessions (while the Friend is currently dropped in) by 30-second increments by tapping a button on the temple of the smartglasses, and they can do this as many times as they want.

\end{design}

\section{ARcall}

\begin{figure}
    \centering
    \includegraphics[scale=1]{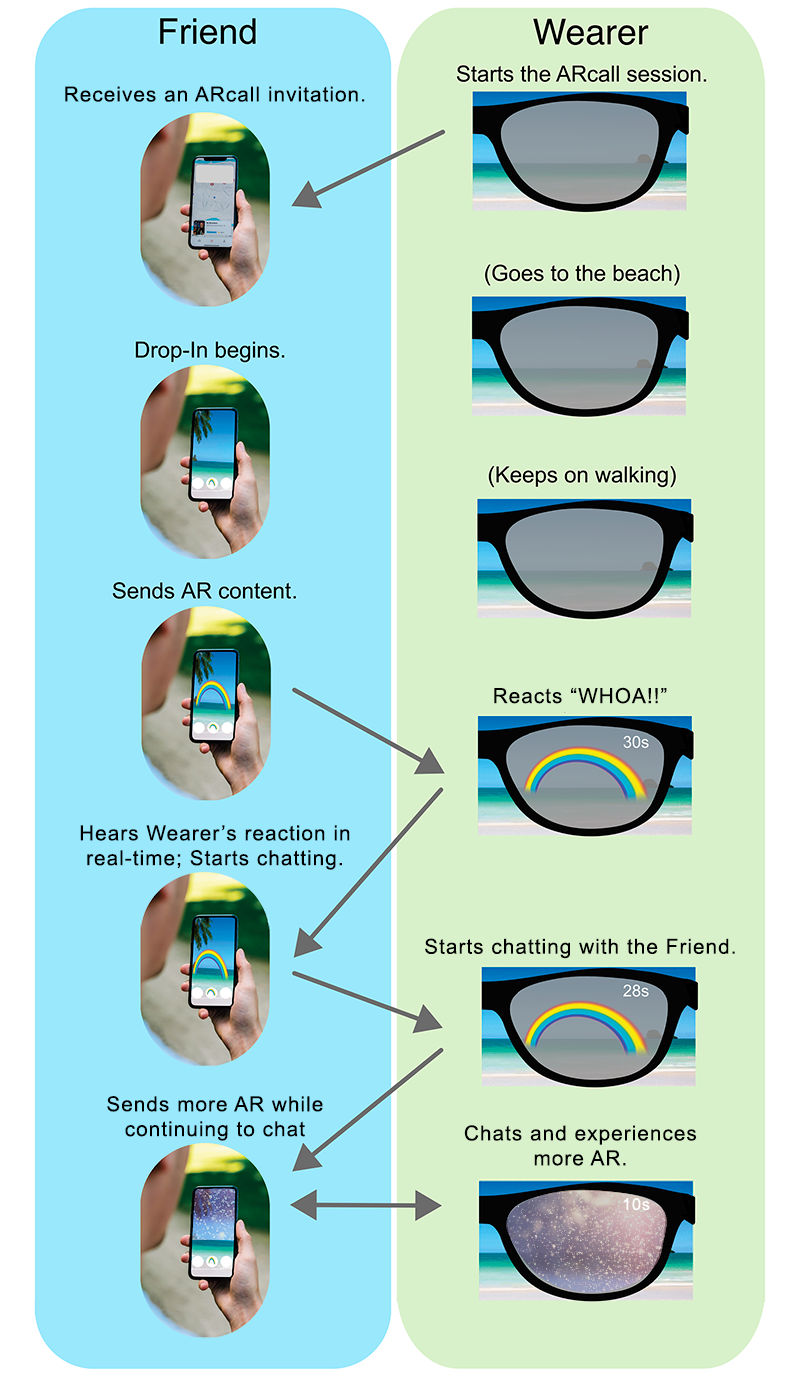}
    \caption{ARcall interaction sequence. The Wearer starts the ARcall session, which invites the Friend to Drop-In. Later, the Friend drops in and surprises the Wearer with an AR rainbow. The Wearer reacts by saying ``WHOA!!'', and the two start chatting while the Friend sends even more AR content.}
    \label{fig:sequence}
    
\end{figure}

\begin{figure}
    \centering
    \includegraphics[scale=0.4]{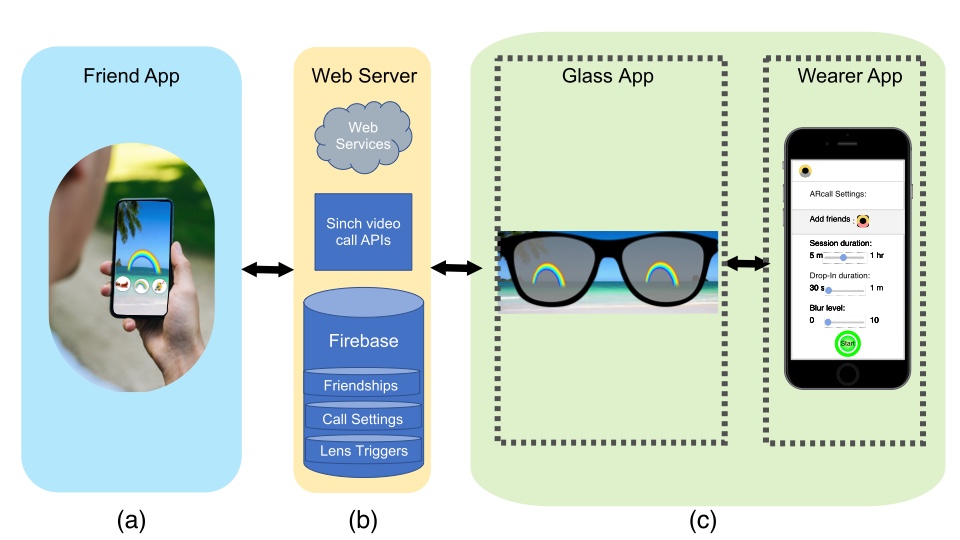}
    \caption{ARcall System Diagram: (a) The \oneq{Friend App} enables a Friend to drop in, send AR content, and talk to the Wearer; (b) The Web server handles video streaming, Friendship database, ARcall preferences, and connects the Wearer with the Friend remotely in real time; (c) The \oneq{Wearer App} lets the Wearer control their preferences and start ARcall sessions. The \oneq{Glass App} enables the Wearer to experience projected AR content.}
    \label{fig:system}
\end{figure}

ARcall is an mixed-device calling system that facilitates AR-based real-time calls between smartphones and smartglasses. It comprises a smartglasses app (Glass App) for the Wearer and a smartphone app (Friend App) for the Friend (see Figure \ref{fig:system}). The Wearer also has a companion smartphone app (Wearer App) to start and end ARcall sessions and configure other settings. This is in line with most consumer camera glasses~\cite{VuzixBlade, focalssetup, SnapSpects}, which have companion smartphone apps to control their settings.

First, the Wearer starts an \textit{ARcall session} to invite their Friend to drop in. The ARcall sessions can last for up to an hour and expires automatically. During this time, the Friend can drop in anytime, initiating \textit{Drop-In sessions}. Drop-In sessions are moments of co-presence between the Wearer and Friend that last for up to a minute each. During these Drop-In sessions, the Friend can engage with the Wearer using AR and voice in real-time. The end-to-end latency is around 100ms.

As an example of a typical interaction, the Wearer, who is going to the beach, might create an ARcall session to invite a close friend to drop in. The Friend can then drop in anytime in the next hour. This allows the Friend to learn the Wearer\rq s context --- where they are and what are they doing --- before sending AR content and interacting with the Wearer. The AR content augments the Wearer's reality. The Friend can experience their reaction, and the two can talk to each other in real time. If the Wearer wants, they can tap on their smartglasses to extend the Drop-In session by 30-second increments, or the call ends, and the Friend can drop in again anytime during the ARcall session. These short Drop-In sessions preserve the battery in the smartglasses, but give the Wearer control to engage with the Friend for as long as they want. ARcall makes it possible for the smartglasses Wearer and their Friend to share a native AR-based experience in real-time.

\vspace{0.5em}
\textbf{Wearer Side:}

Our \textit{Wearer app} is written in Swift and runs on iOS. The Wearer app runs on a companion smartphone. This app provides controls to set the duration of an ARcall session, to start and end the session, and to select a Friend to send the ARcall invite to. The Wearer sets the ARcall parameters before inviting their Friend. These parameters are then pushed to a cloud database server, Firebase \cite{firebase}.

Our \textit{Glass app}, which runs on the Wearer's smartglasses, is written in Java and runs on Android operating system version 5.0 and API level 21. ARcall was implemented on Snap's 4th Generation AR Spectacles (2021) \cite{Spectacles}. We used the Sinch video calling API \cite{sinch} for both the Glass app and the smartphone app used by the Friend. The Glass app pulls the ARcall session parameters from the cloud database server. Note that these parameters are set in the Wearer app before starting the ARcall session. The Glass app also listens to the taps on the right temple of the smartglasses.

\vspace{0.5em}
\textbf{Friend Side:}

\textit{Friend App} is written in Swift and runs on iOS. The Wearer app runs on a smartphone. This application used ARKit \cite{arkit}, SceneKit \cite{scenekit}, and SpriteKit \cite{spritekit} to render AR content on the smartphone. It also employs the Sinch video calling APIs \cite{sinch} to support high-quality video calls with the Wearer. This application supports three tasks: scrolling through the list of AR content, sending the AR content to the Glass App, and the ability to mute microphone audio.

\section{Main Design Components}

We introduce three main components of ARcall, namely \textit{Drop-In}, \textit{ARaction}, and \textit{Micro-Chat}. We describe these components along with the corresponding application features available to both the Wearer and the Friend.

\subsection{Drop-In}
 Drop-In allows the Friend to see the Wearer's live context --- where they are and what they are doing --- by \quotes{dropping in}and seeing a video feed of the Wearer’s point of view (see Figure \ref{fig:teaser}~(a)). This is an invite-only feature between the Wearer and the Friend.

% \vspace{0.5em}
\textbf{Wearer Side:}

\textit{Inviting the Friend to the ARcall session (Wearer App)}: The Wearer specifies the Friend they are inviting before beginning the ARcall session. The information about the Friend is maintained in the cloud database server, Firebase \cite{firebase}. The Wearer can add or remove the Friend as desired.

\textit{Setting the blur level (Wearer App)}: The Wearer can control the way their point of view (POV) is shared with the Friend. The blur level can be set to a value from 0 (high quality video stream) to 10 (completely blurred video stream). Blur control gives the Wearer a flexible way to maintain their privacy and to control their visual feed prior to inviting the Friend to drop in (\autoref{design:privacy}).

\textit{Setting an ARcall session duration (Wearer App)}: The ARcall session's duration (i.e., the period during which the Friend is allowed to drop in) can be set for 5 minutes to an hour. The Wearer has to explicitly tap the \quotes{start} button to begin the ARcall session.

\textit{Setting a Drop-In session duration (Wearer App)}: The Wearer can also set how long each Drop-In session (period of co-presence with the Friend) can last within the greater ARcall session. We call this the Drop-In session duration. We conducted empirical tests to assess how long the smartglasses could support video streaming before becoming too hot (\autoref{design:power}). Informed by this test, we restricted the possible duration to between 30 seconds to a minute. Regardless of the chosen duration, the Wearer can extend an active Drop-In session (one in which their Friend is currently dropped in) by 30-second increments, and they can do this as many times as they want.

\begin{figure}[h]
    \centering
    \includegraphics[scale=0.35]{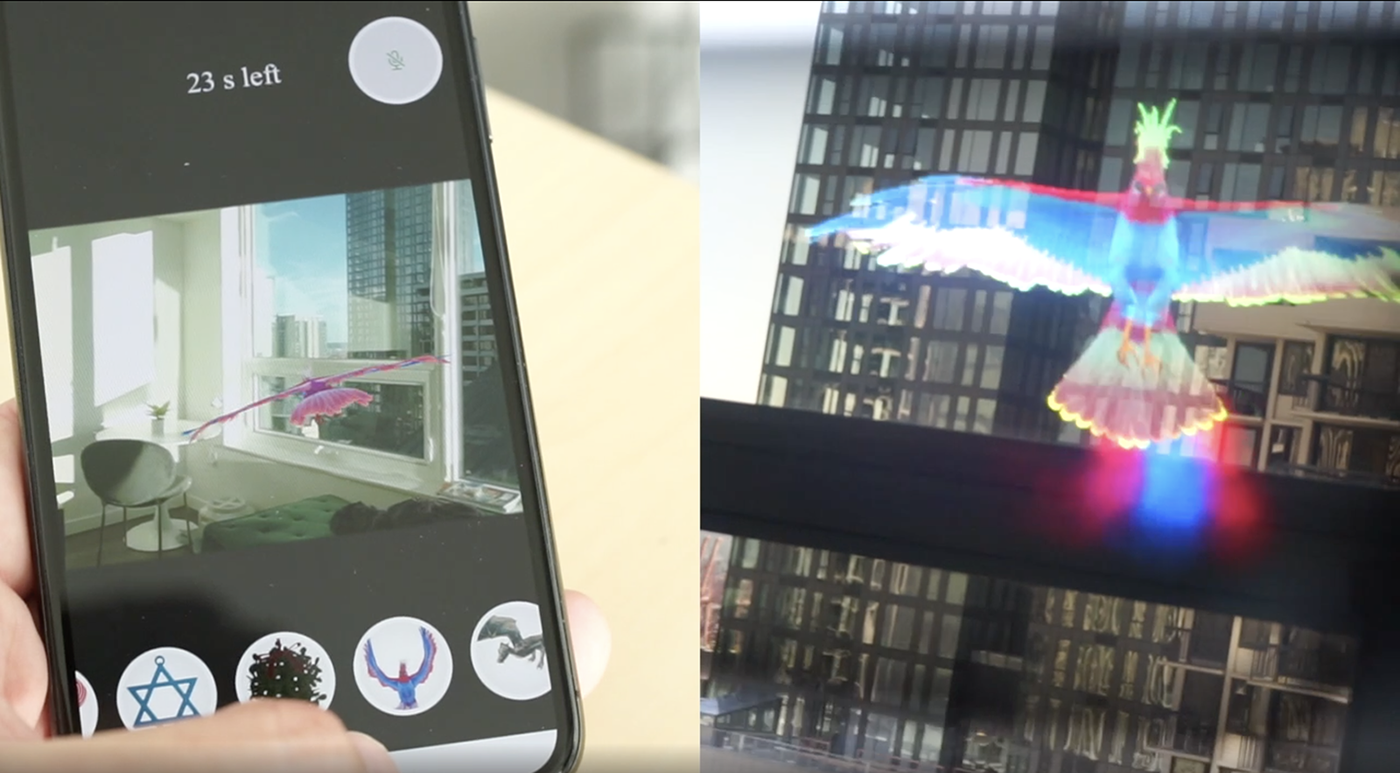}
    \caption{(Left) The Friend selects an AR phoenix from a carousel on the Friend App. (Right) The AR phoenix is immediately projected onto the Wearer's point of view.}
    \label{fig:snapshot}
    \vspace{-3mm}
\end{figure}

% \vspace{0.5em}
\textbf{Friend Side:}

\textit{Reacting to an invitation (Friend App)}: When the Wearer begins the ARcall session, the Friend will receive a Drop-In invitation on their smartphone. The Friend can then drop in anytime by tapping on the invitation.

\textit{Seeing the Wearer's POV (Friend App)}: When the Friend participates in the Drop-In session, they can see and hear what's happening in the Wearer\rq s environment via a camera feed. This camera feed is placed at the center of the smartphone display, as shown in Figure \ref{fig:snapshot}. The remote Friend can gauge the Wearer's context --- where they are and what are they doing --- in order to send them AR content (\autoref{design:interrupt}).

\subsection{ARaction}
ARaction component enables interaction with the Wearer using AR content.
%ARaction allows the Friend to communicate with the Wearer natively in AR using purely AR-based content (\autoref{design:interactivity}). 
The Friend augments the Wearer\rq s reality by projecting AR content onto their POV. %(\autoref{design:interactivity}). 
As illustrated by Figures \ref{fig:sequence} and \ref{fig:snapshot}, the Friend can send AR content in real-time after dropping in.

% \textbf{Wearer Side:}
% \textit{Always ready to project anytime. Focus on native AR narration.}

% \vspace{0.5em}
\textbf{Friend Side:}

\textit{Selecting the AR Content (Friend App)}: We carefully selected AR content that fits the display and can be seen clearly by the Wearer. With \autoref{design:fov} and \autoref{design:lenses} in mind, we chose two types of AR content. The first type consists of \textit{animated 3D virtual objects}, such as a flying bird or a dragon, that are placed in the middle of the smartglass display when sent by the Friend (see Figure \ref{fig:ARcontent}). Animated content can improve the quality of expression while conversing \cite{an2021vibemoji}. These 3D objects will follow the Wearer\rq s head position and will remain visible, staying in the field-of-view of the smartglasses. The second AR content type consists of \textit{particle objects}, such as snow flakes, that completely cover the smartglass display. In total, we employed ten pieces of AR content. The AR content was organized as a horizontal carousel at the bottom of the screen (see Figure \ref{fig:snapshot}), similar to popular photo- and video-sharing apps such as Snapchat \cite{snapfacefilter} and Instagram \cite{insta}. Only one piece of AR content can be displayed to the Wearer at a time, but the Friend can send as many pieces of AR content as they wish.

\textit{Sending AR content (Friend App)}: After the Friend drops in to gauge the Wearer's context, they can choose a piece of AR content to project to the Wearer\rq s view. While Drop-In lets the remote Friend \textit{peek} into the Wearer\rq s context, ARaction lets them \textit{project} AR content at appropriate times (\autoref{design:interrupt}).

\subsection{Micro-Chat}
A Micro-Chat is a time-bounded voice call between the Wearer and the Friend that happens alongside the ARaction and adds meaning to it. It starts as soon as the the Drop-In session starts, specifically, once the Friend drops in to see the Wearer's point of view. 

% \vspace{0.5em}
\textbf{Wearer Side:}

\textit{Extending a Drop-In session (Glass App)}: A countdown timer shows the time remaining in the Drop-In session. The timer's state is synced with the Glass App. The Wearer can extend the Drop-In session by 30-second increments during the session by pressing a button located on the right temple of the smartglasses.

% \vspace{0.5em}
\textbf{Friend Side:}

\textit{Setting audio controls (Friend App)}: %Audio controls lets the remote Friend decide when they would like to start talking with the Wearer. %(\autoref{design:interactivity}). 
The Friend's smartphone app lets them \oneq{mute} or \oneq{unmute} their microphone, so they can decide when they would like to start talking to the Wearer (see Figure \ref{fig:snapshot} (left)).

\section{FIRST-USE STUDY} \label{ResearchQuestions}
\begin{comment}
What we want to evaluate? why?
\end{comment}

This study qualitatively evaluated the following four aspects of the ARcall: (1) the overall user experience, (2) initial perceptions of using ARcall, (3) the feeling of connectedness, and (4) the friction points of using ARcall.

\subsection{Participants}
In total, 14 volunteers participated in the study. Our study group consisted of four females and ten males, with two participants in their 20s and 12 in 30s. The participants had diverse backgrounds and included a 3D artist, a product designer, a hardware engineer, and a software engineer. We asked participants to participate in pairs with their friends, so in total, we had seven sessions. Six participants had prior experience with smartglasses, seven participants had experience using AR filters, and nine participants were used to communicating with their friends using photos or videos using a smartphone. All Wearer participants had normal or corrected vision and did not wear prescription glasses.

\subsection{Procedure}

The participants went through a five-step procedure: a demographic questionnaire, ARcall onboarding, ARcall experience testing, a post-experience questionnaire, and a semi-structured interview. The ARcall experience testing took 10-15 minutes, and the study took roughly 60 minutes.

\begin{figure}
    \centering
    \includegraphics[width=0.22\textwidth]{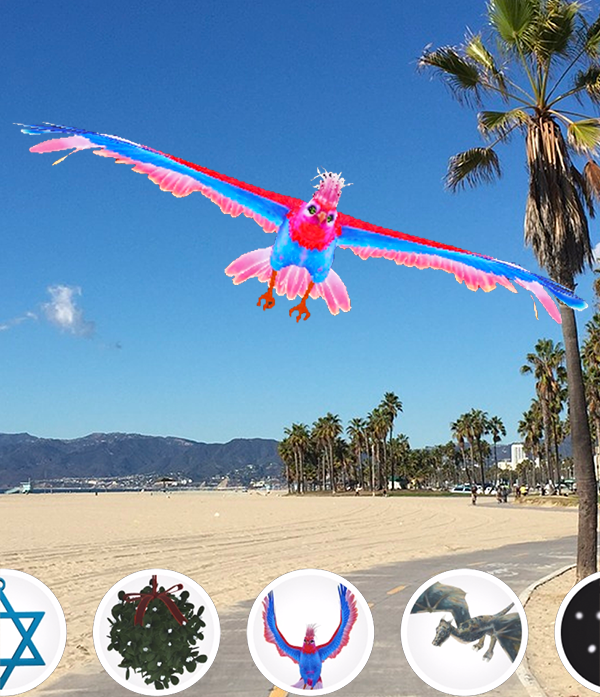}
    \includegraphics[width=0.22\textwidth]{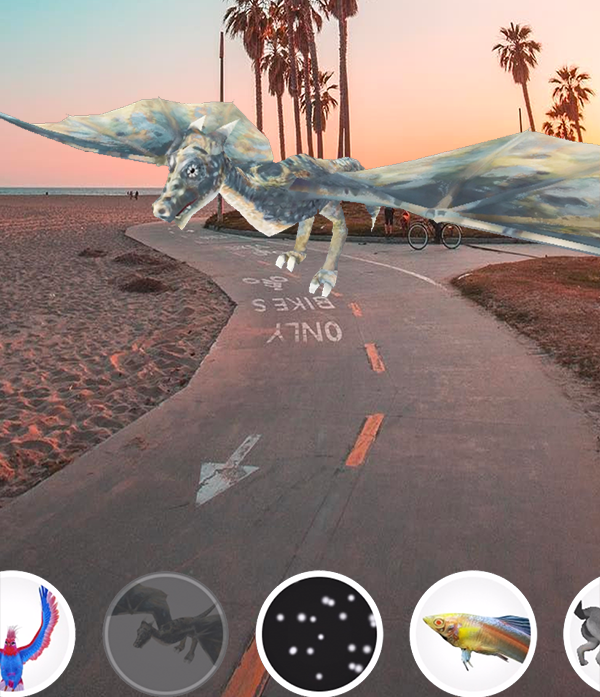}
    \includegraphics[width=0.22\textwidth]{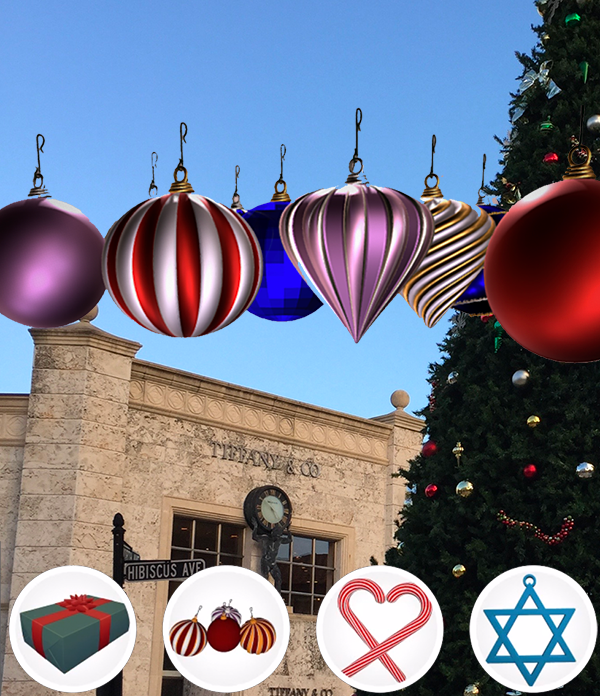}
    \includegraphics[width=0.22\textwidth]{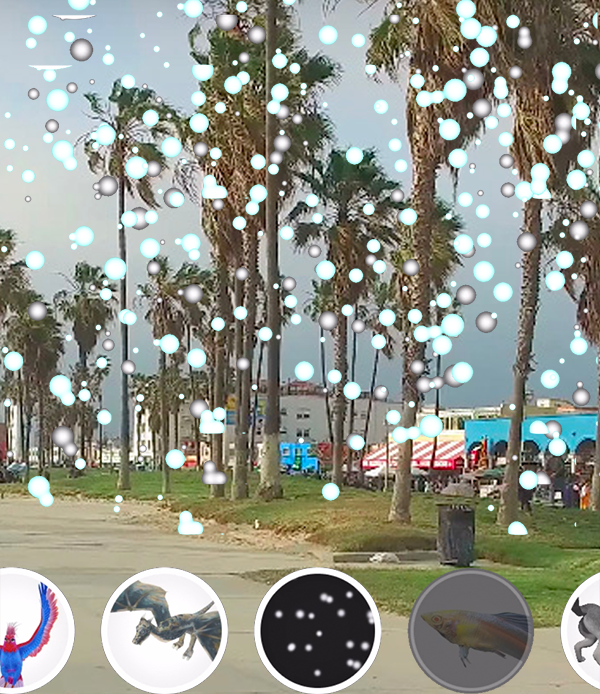}
    \caption{Sample AR content for the Friend. We show (left to right) a phoenix, a dragon, holiday ornaments, and snow.}
    \label{fig:ARcontent}
\end{figure}

First, each pair of participants completed a demographic survey that asked them to indicate their age, gender, and relationship. Next, we conducted a 20-minute onboarding session in which we showed the participant pairs how to use ARcall and explained the available AR content. This try-it-yourself onboarding session ensured that we could mitigate the novelty effect of using smartglass. Since we conducted our investigation during the holiday season, we themed the AR content around holiday festivities and fantasy.

Next, for a portion of the ARcall experience testing, we asked the pairs to imagine that they lived in different places. We asked the Friend participant to use ARcall to send the Wearer holiday cheer, and we asked the Wearer to take a walk outside and invite the Friend to drop in. The Friend stayed in a conference room while the Wearer took a walk outside. Over 70 percent of informal calls in daily life consist of small talk and showing people things from our immediate environments that we would like to discuss \cite{o2006everyday}. These informal calls are typically short in duration, spanning 10-15 minutes each. Considering the primary use of camera glasses when wearers are engaged in outdoor activities \cite{bipat2019analyzing, neustaedter2012moving} and the limited cognitive resources at the Wearer's disposal \cite{oulasvirta2005interaction}, the ARcall experience was designed to run for 10-15 minutes.

\begin{figure}
    \centering
    \includegraphics[scale=0.5]{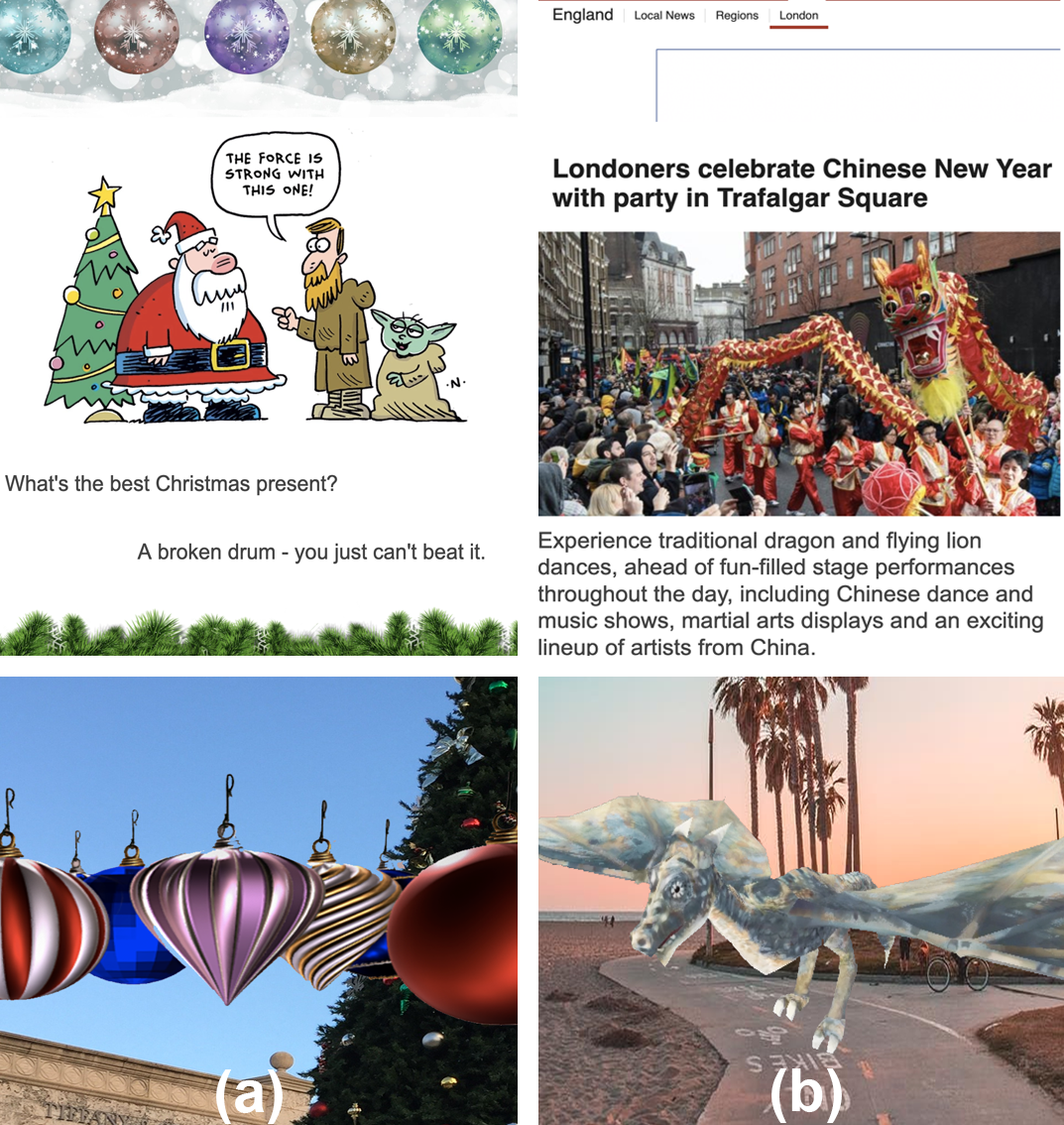}
    \caption{Sample visual artifacts and corresponding AR content. (a) A Santa surrounded by ornaments and AR content of holiday ornaments; (b) A Dragon Festival news excerpt and AR content of a dragon.}
    \label{fig:posters}
\end{figure}

All of the AR content available for the Friend to send was festivity or fantasy themed (see Figure \ref{fig:ARcontent}), as the study took place around the holiday season. The available content included a bird, a dragon, holiday ornaments, holiday gifts, snow, a Star of David, a whale, a mistletoe sprig, a unicorn, a reindeer, and a Santa on a sleigh. To observe whether the Wearer\rq s context played a role in the AR content selected, we placed six visual artifacts along the Wearer's walking path. Each artifact was themed to subtly correspond to a piece of AR content that the Friends could select. For example, a news story poster with a banner image of a dragon. Figure \ref{fig:posters} (b) top highlights a sample of the visual artifacts that we placed on the walking path, along with their associated AR content.

Next, we asked the pair to complete a post-experience questionnaire that probed the users' experience with Drop-In, ARaction and Micro-Chat, which included open-ended questions with responses on a 7-point Likert scale (i.e., \quotes{1} indicated Strongly Disagree and \quotes{7} indicated Strongly Agree). Lastly, we conducted a semi-structured interview to capture remarks that the questionnaire may not have elicited.

\subsubsection{Data collection and Analysis}

The software application logs were analyzed to derive quantitative metrics like the median duration of the Drop-In session, the median number of AR content items sent, and the total number of times the session was extended. Further, a thematic analysis of the post-experiment responses and the semi-structured interview responses was conducted. Two separate researchers iteratively coded the collected data, then merged their findings to identify the common themes that emerged on further analysis. We discuss these themes for each ARcall design component in the next section. Our evaluation protocol and analysis method are similar to Judge ~\ea's investigation \cite{judge2010sharing}.

\section{RESULTS}

The median Drop-In session duration was 1.9 minutes, and the median number of pieces of AR content sent per Drop-In session was 11. Over 200 pieces of AR content were sent in total, and Wearers extended the duration of the Drop-In sessions 48 times. Overall, the participants provided positive feedback on the experience of using ARcall and felt that it would add great value to smartglasses (\median{6}). 

\subsection{Experience with ARaction} \label{results:araction}

Wearers reported that the AR content they got from Friends was highly relevant to their context (\median{6}) and felt that it augmented their reality (\median{6}). Moreover, they were able to understand Friends\rq intent behind the AR content (\median{6}). This shows that ARcall's three components worked together effectively, allowing Friends to drop in to gauge the Wearer's context, send AR content relevant to that context, and use their voices to convey their intent --- all simultaneously, in real-time.

\vspace{-1mm}
\subsubsection{Both Wearers and Friends found that ARaction could enable moments of surprise}

On the Wearer side, participants were often pleasantly surprised by the AR content. For instance, W13 noted, \usersaid{I was surprised [and] I felt closer to [the Friend.]} Also, W9 added, \usersaid{[ARaction] adds an element of surprise, which is really cool to see[.]} 

On the Friend side, projecting the AR content and getting the Wearer's reaction was perceived positively, and seeing the Wearer's reaction acted as a key motivator for sending them content. Friends felt very comfortable sending as much AR content as they wished (\median{6}) and did not feel hesitant about sending AR content (\median{6}), likely as a result of being specifically invited to the ARcall session by the Wearer. F10 commented, \usersaid{[ARaction was] unique because of the ability to shock people with the AR content.}

\vspace{-1mm}
\subsubsection{Both Wearers and Friends felt connected and found ARaction fun to use}

On the Wearer side, W7 observed, \usersaid{Adding the AR content makes the interactions fun. I can see this being a great way to have fun with friends without having the event be too formal. Simply messing around with AR content [...] from a Friend [while I'm on a stroll] makes me feel special that someone reached out to me.} W3 added, \usersaid{They add somewhat of a gamification feel to my world which was [...] unique, and I wanted to try it out more and with more friends.} This participant went on to add, \usersaid{I think it's a superior experience [compared to livestreaming and video calls]. Life suddenly feels gamified, and you have your friend with you --- there's more opportunities to share [...]. And the fact that you are not looking at your phone, you are [freer] to explore [the world around you].}

Friends felt that getting the Wearer's reaction was highly important to the AR content they projected (\median{6}), and they had fun seeing the Wearer\rq s reactions firsthand (\median{6}). Friends also felt that hearing the Wearer heightened their sense of togetherness (\median{6}). F16 remarked, \usersaid{It was an awesome way to have fun with my friend and mess with them. :) Sending AR content jokingly or situationally was interesting.} 

\vspace{-1mm}
\subsubsection{Both Wearers and Friends found ARaction to be personalized and expressive}

The Friend's ability to gauge the Wearer's context before projecting AR content, including their ability to pick up on the subtle visual artifacts that we put in place, made Wearers feel that Friends had hand-picked the AR content for them. For instance, W3 observed, \usersaid{Based on what I was looking at, they augmented the AR content to my reality, and it was pretty spot on.} W6 pointed out the way the context was utilized by their Friend: \usersaid{Pretty neat to have telepresence with another, and for them to be able to interact with what I was seeing.} 

From the Friend's perspective, F4 felt, \usersaid{[This whole] experience make[s] our conversation [...] closer to each other, and it will blur the boarders and on more step[s] to brain-to-brain conversation.}

\vspace{-1mm}
\subsubsection{Both Wearers and Friends found ARaction to be immersive}

ARaction enabled the AR content to be immersive for both parties. W3 noted, \usersaid{[I was s]urprised. I felt like my friend was somewhere in the heavens looking at me. [...] I think it did feel like my friend augmented my reality.}

From the Friend's perspective, Friends felt that they had the power to change the Wearer's reality. F12 said, \usersaid{With the snow I was able to make [W12] feel like he was in a different place, which to me is definitely augmenting his reality.} Moreover, F12 highlighted the value of ARaction: \usersaid{I think it adds quite a lot because it's more than just talking. It makes you feel like you are almost experiencing the same thing as the person.} F12 continued, \usersaid{I think it is a benefit because it makes it a shared experience for the two of you, instead of one person joining another person's experience.}

\vspace{-1mm}
\subsubsection{Friends wished to be offered a wider selection of AR content} \label{results:morelenses}

Our participants had ten pieces of AR content to choose from, but they felt they needed more. W11 said that, for them, compelling AR meant \usersaid{anything that is relevant or contextual to the environment or feels more blended with the reality in front of me,} strengthening the case for a wider selection of AR content. Moreover, studies have shown that a series of multiple emoticons can help convey new meanings \cite{khandekar2019opico} and, at times, convey meanings that go beyond the original intended meanings \cite{kelly2015characterising}. W5 noted, \usersaid{The AR content w[as] decorative, but [it] didn't necessarily capture how I was feeling.} F14 suggested additions such as \usersaid{Dog, cat, cars, [and] real world objects like tree[s].} W5 suggested \usersaid{ones that feel relevant to the current view, my feelings, or my friend's feelings.} This suggests a need to develop more sophisticated AR content that can convey an understanding of a scene and be tailored to users' emotions. 

\vspace{-1mm}
\subsubsection{Both Wearers and Friends wished to control the placement of AR content} \label{results:lossofcontrol}

% \subsubsection{Friend felt to have more control over the presentation of the contents}
Another direction for improving the experience is letting the Wearers move the AR content that is projected. Our current selection of AR content consisted of either particle objects (e.g., snow) or 3D virtual objects (e.g., a bird) that were automatically positioned and could not be moved. For instance, a projected bird would always fly in the center of the screen, and the snow would always take over the screen to simulate a snowfall.  W11 lamented, \usersaid{I did feel a loss of control. Once the AR content appeared, it took over a lot of my field of vision, so I was no longer able to read the [news story that I was reading].} 

Friends acknowledged the issue of occlusion as well. F10 commented, \usersaid{It was fun to surprise them with different AR content although I felt like I was blocking their view the whole time.} 

\vspace{-1mm}
\subsection{Experience with Drop-In}
Most of the Wearers set the session duration to the maximum value of one minute and the camera feed to the \quotes{No Blur} setting. We suspect that the blurring felt unnecessary because the participants trusted each other during the study. Yet, the blurring feature might be necessary for a different usage context \cite{lederer2002conceptual, denning2014situ} or if the degree of trust between the Friend and the Wearer varies \cite{kyto2021strangers, jones2011contextual}. The Wearers indicated that ARcall's Drop-In feature protected their privacy (\median{6}) and that they were also comfortable sharing their POV with their chosen Friend (\median{6}). The ability to extend or end the call at any time is highly important to the Wearers (\median{6}).

When they considered design aspects \autoref{design:interrupt} and \autoref{design:privacy}, the Friends echoed these sentiments in their responses to the following prompts: \quotes{Since my friend [the Wearer] invited me to their ARcall session, I did not feel that I was invading their privacy} (\median{6}) and \quotes{I felt comfortable seeing the Wearer's POV} (\median{6}). Drop-In elevated the sense of togetherness between the two parties (\median{6}). W15 was skeptical initially, setting the blur option to \quotes{Full blur}; however, this participant later changed the setting to \quotes{No Blur} once they began experiencing ARcall.

\vspace{-1mm}
\subsubsection{Drop-In not only helped Friends determine the Wearer's context, it made them feel a sense of togetherness in relation to the Wearer}
Our semi-structured interview feedback shed light on the value of Drop-In. W3 felt that \usersaid{Drop in was a very interesting experience. I felt [like I was] being guided by my friend and interestingly felt closer [to them] in that moment.} W9 stated, \usersaid{It is really cool that I can allow a friend to drop in because I am waiting for [the] surprise of them joining.} W7 felt that \usersaid{[it] makes it fun that friends can pop in and have fun with your worldview.} W3 echoed this sentiment: \usersaid{I think the Drop-In adds [to] the augmented reality part of the experience. I feel like it's very important to see what the Wearer is seeing.} Friends also felt that Drop-In was necessary for effective AR. F8 remarked, \usersaid{I think this adds a lot of value [to AR]. I really like[d] being able to see what my friend was seeing.} F4 highlighted the convenience associated with the ability to drop in: \usersaid{[I]t's fun and it [was a] really convenient way to show me what [W4] [was feeling and seeing] at that moment.} F10 and F4 observed, \usersaid{I think it's a unique experience seeing through the eyes of the other person and being able to interact, not face-to-face, but through their eyes.}

\vspace{-1mm}

\subsubsection{Drop-In did not seem to create privacy concerns, but Wearers wanted an indicator that showed when their Friend was present}\label{results:morecontext}

The Wearers strongly agreed that Drop-In's invite-only nature was an important feature for controlling their privacy (\median{7}). This supports design considerations \autoref{design:interrupt} and \autoref{design:privacy}, which recommend that allowing the Wearer to invite only one Friend would make ARcall a trustworthy and safe experience. In our ARcall implementation, there was no indication of when a Friend had dropped in; we designed it that way to make it more likely for the Wearer to be surprised by the Friend's AR in a fun way. This setup worked great for most of the Wearers, but some expressed the desire for an indicator. W7, for example, expressed the need for \usersaid{a better way to know when [F7] starts watching what I am sharing,} adding that, \usersaid{[...] unless my friend [projected] an AR experience, I couldn't tell if they had joined.}

\vspace{-1mm}
\subsection{Experience with Micro-Chat} \label{results:microchat}

Wearers found several aspects of Micro-Chat highly engaging. They responded positively to the following prompts: \quotes{I felt together with my Friend as soon as I heard them talk} (\median{6}) and \quotes{It was fun to [Micro-Chat] with my friend within my chosen time duration} (\median{6}). On the Friend side, participants felt it important to chat with the Wearer right after projecting AR content to them (\median{6}). They also found it fun to experience the Wearer\rq s reactions using Micro-Chat (\median{6}).

\vspace{-1mm}
\subsubsection{Both Wearers and Friends found that Micro-Chat added intimacy}

Wearers felt that Micro-Chat provided instant context to their experience. W9 commented, \usersaid{It was my favorite thing---hearing my friend right away and knowing that they [could] see what I [projected].} W3 said, \usersaid{I think more than the AR experience, the voice chat was fantastic. I think this feature really builds [on] the human factor of [smartglasses]: the communication side.} 
For Friends, experiencing the Wearer\rq s reaction to their AR content gave the Friend a sense of being together with the Wearer (\median{6}). Friends found it important to get the Wearer's feedback on their chosen AR content (\median{6}) using Micro-Chat. F6 remarked, \usersaid{[Micro-Chat] adds context to the AR content.} F16 asserted, \usersaid{ARcall would be nothing without its interaction aspects. Voice and AR content are the only way to interact with [a friend wearing AR glasses]. Critical.} 

\subsubsection{Friends felt that the duration of Micro-chat was short} \label{results:lengthymicrochat}

There were a few exceptions where the Friends felt Micro-chat was limiting. F14 stated, \usersaid{Most video calls are longer for me, and I would prefer anything more than a sound byte.} We also saw contrary remarks, for instance, F6 pointed out, \quotes{I can see it being helpful for quick questions and fast interactions.} Specific to F14\rq s concern about the call length, ARcall does provide the wearer with a way to extend the call. So, in a way, it depends on the the mutual decision of the parties in the ARcall experience. The Wearers indicated no concerns regarding the Micro-chat.

\section{Design Implications and Limitations}

\designguide{Drop-Ins enable surprise, but Wearers should be able to know about Friends' presence as soon as they drop in:}
\noindent
While most Wearers enjoyed being surprised by AR content, some wanted more context about their Friend's presence (see \ref{results:morecontext}). In future AR communication apps, Friends' information can be presented more explicitly along with the projected AR content (e.g., ``John just dropped-in!''). %were wary of using it throughout the day. 
There may be times when a surprise is undesirable, and the wearer may benefit from increased control over incoming call requests. For example, for non-subtle AR lenses, it would be useful to snooze the incoming call or have the option of seeing a basic LED-like notification when AR content arrives rather than the content itself.

% \subsection{Design Implications from ARaction}

% \vspace{-1mm}
\designguide{AR communications systems should feature a wide variety of AR content and intelligent scene-based recommendations:}
\noindent
%Recall in our user study, we ask the participants to imagine a holiday season, essentially, directing the imagination of the participants. Also, 
Results (\ref{results:morelenses}) show that Wearers felt that AR content was highly personalized when they noticed its relevance to a visual artifact that they were looking at. %For example, when they were looking atSince the wearer knows that the remote friend is looking at their POV when they drop in. So, when they receive the AR content, they interpret these contents based on where they are currently looking at. 
Thus, future AR communication systems should provide a gallery of AR content for Friends to search through so that they can match the AR content they send with the Wearer's situation. Alternatively, the app could be designed as an AI-based system that can scan the Wearer's scene and recommend relevant AR content to the Friend. For instance, the Friend App could recommend a beach ball or a sandcastle if the Wearer were at the beach. However, designers should carefully craft such experiences, balancing automation and freedom of expression \cite{an2021vibemoji}.

% \vspace{-1mm}
\designguide{AR content should be chosen to minimize the difference between the way Friends see the content and the way Wearers see it:}
% %\designguide{AR content should be chosen to minimize perception differences between the Wearer and the Frienddesigners should be thoughtful about how they display the AR content  Wearer's view to the Friend:}
\noindent
We observed that Friends' perception of how the Wearer would see the AR content they sent did not always match the way the Wearer actually saw it (see \ref{results:araction}).
% Based on the subjective feedback, we observed the differences between the way some AR content was perceived by the Wearer and the remote friend. 
For instance, a Friend mentioned \usersaid{With the snow AR content, I was able to make [the Wearer] feel like he was in a different place, which to me is definitely augmenting his reality.} However, the Wearer observed that \usersaid{[..] some content like the snow and the orcas [felt] strange [because they were not] full-screen.} This difference in perception between the Friend and Wearer occurs whenever the AR content is displayed edge-to-edge on the Friend's smartphone screen. It was detrimental to the Wearer\rq s experience. They felt that the display became obvious and that it disrupted their sense of immersion. In the future, designers should take into account such rendering differences when designing for AR communication.

\designguide{AR content should be movable and anchored to real world objects and surfaces:} 
Results (\ref{results:lossofcontrol}) revealed two important principles for how AR content should be presented to smartglass wearers. First, the Wearers consider AR more convincing when it integrates with their environment realistically. For example, a palm tree model should be pinned to the ground and a bee model should be floating in the air, in the same way they would appear in the real world. The concept of native AR communication centers on immersion, so future AR communication designers should aim to anchor AR content to real-world objects or surfaces to make it more convincing. Second, both Wearers and Friends found that AR content sometimes occluded their view, so future AR communication systems should allow them to control the position of AR content within the environment.

\designguide{The short duration of Micro-Chats should be seen as a means of enabling new social experiences rather than simply prolonging battery life:}
% We observed that the wearer would often extend the Micro-Chat multiple times to have a longer conversation with the friend, very often exceeding a minute in duration. Hence, to ensure a satisfying ARcall experience, Micro-Chats should be at least a minute in duration by default.
While we put a limit on the duration of Drop-In sessions to preserve the battery life of the smartglasses, this limit offers a valuable benefit (see \ref{results:microchat}): it can enable low-commitment, short bursts of interactions between friends, similar to a ``hallway chat.'' The Drop-In concept, together with Micro-Chat, has the potential to create an "office-hour"-like experience in which, once the Wearer indicates their availability (i.e., ``props open their office door''), their Friend can drop in and have a brief interaction before moving on to their business. Studies have found that short-lived interactions, even with strangers, can induce positive feelings \cite{epley2014mistakenly}. While we did not investigate these, there are additional ways to reduce power consumption, such as lowering display brightness in low-light situations or using only one side of the display. Future designers can explore this space further and use these new design components to create compelling experiences, including components that involve more than one Friend or allow functionality when bandwidth is limited \cite{kaye2006just}.

\smallbreak
\noindent
Although the proof-of-concept system demonstrated in this research could be improved further, for instance, by adding scene understanding to stylize AR lenses to the Wearer's surroundings, we feel that pursuing these improvements is a promising avenue for future research. Furthermore, because cameras are becoming more powerful, future systems could employ vision-based techniques, without bothering the user \cite{xu2015exploring}, to intelligently situate the AR content in order to offer a more contextual AR experience. We hope the design implications of ARcall will help designers craft novel AR communication methods and allow researchers to investigate further improvements.

\section{CONCLUSION}

The idea of using AR as a core medium of communication is an essential step in pushing the boundaries of technology-mediated communication. ARcall is a new type of calling system that explores native AR-based communication. Our investigation with 14 participants found that ARaction can enable moments of immersion, playfulness, surprise, and fun. We found that Drop-In helped Friends gauge Wearers’ real-time context and allowed them to personalize the ARaction for the Wearer. Additionally, Micro-Chat creates intimacy and makes the moments of co-presence more effective. Our findings about user behaviour and the design insights derived from this study can inspire designers and researchers to explore a new design space for native AR communication systems. We hope AR-based communication systems like ARcall can enable people to express themselves in ways that were not possible before.

%%
%% The next two lines define the bibliography style to be used, and
%% the bibliography file.
\bibliographystyle{ACM-Reference-Format}
\bibliography{_main}

%%% -*-BibTeX-*-
%%% Do NOT edit. File created by BibTeX with style
%%% ACM-Reference-Format-Journals [18-Jan-2012].

\begin{thebibliography}{80}

%%% ====================================================================
%%% NOTE TO THE USER: you can override these defaults by providing
%%% customized versions of any of these macros before the \bibliography
%%% command.  Each of them MUST provide its own final punctuation,
%%% except for \shownote{}, \showDOI{}, and \showURL{}.  The latter two
%%% do not use final punctuation, in order to avoid confusing it with
%%% the Web address.
%%%
%%% To suppress output of a particular field, define its macro to expand
%%% to an empty string, or better, \unskip, like this:
%%%
%%% \newcommand{\showDOI}[1]{\unskip}   % LaTeX syntax
%%%
%%% \def \showDOI #1{\unskip}           % plain TeX syntax
%%%
%%% ====================================================================

\ifx \showCODEN    \undefined \def \showCODEN     #1{\unskip}     \fi
\ifx \showDOI      \undefined \def \showDOI       #1{#1}\fi
\ifx \showISBNx    \undefined \def \showISBNx     #1{\unskip}     \fi
\ifx \showISBNxiii \undefined \def \showISBNxiii  #1{\unskip}     \fi
\ifx \showISSN     \undefined \def \showISSN      #1{\unskip}     \fi
\ifx \showLCCN     \undefined \def \showLCCN      #1{\unskip}     \fi
\ifx \shownote     \undefined \def \shownote      #1{#1}          \fi
\ifx \showarticletitle \undefined \def \showarticletitle #1{#1}   \fi
\ifx \showURL      \undefined \def \showURL       {\relax}        \fi
% The following commands are used for tagged output and should be
% invisible to TeX
\providecommand\bibfield[2]{#2}
\providecommand\bibinfo[2]{#2}
\providecommand\natexlab[1]{#1}
\providecommand\showeprint[2][]{arXiv:#2}

\bibitem[\protect\citeauthoryear{Ackerman}{Ackerman}{2000}]%
        {ackerman2000intellectual}
\bibfield{author}{\bibinfo{person}{Mark~S Ackerman}.}
  \bibinfo{year}{2000}\natexlab{}.
\newblock \showarticletitle{The intellectual challenge of CSCW: The gap between
  social requirements and technical feasibility}.
\newblock \bibinfo{journal}{\emph{Human--Computer Interaction}}
  \bibinfo{volume}{15}, \bibinfo{number}{2-3} (\bibinfo{year}{2000}),
  \bibinfo{pages}{179--203}.
\newblock


\bibitem[\protect\citeauthoryear{Ackerman and Mainwaring}{Ackerman and
  Mainwaring}{2005}]%
        {ackerman2005privacy}
\bibfield{author}{\bibinfo{person}{Mark~S Ackerman} {and}
  \bibinfo{person}{Scott~D Mainwaring}.} \bibinfo{year}{2005}\natexlab{}.
\newblock \showarticletitle{Privacy issues and human-computer interaction}.
\newblock \bibinfo{journal}{\emph{Computer}} \bibinfo{volume}{27},
  \bibinfo{number}{5} (\bibinfo{year}{2005}), \bibinfo{pages}{19--26}.
\newblock


\bibitem[\protect\citeauthoryear{{Alpha Exploration Co.}}{{Alpha Exploration
  Co.}}{2020}]%
        {clubhouse}
\bibfield{author}{\bibinfo{person}{{Alpha Exploration Co.}}}
  \bibinfo{year}{2020}\natexlab{}.
\newblock \bibinfo{title}{Clubhouse}.
\newblock
\newblock
\newblock
\shownote{\url{https://www.joinclubhouse.com}.}


\bibitem[\protect\citeauthoryear{Ames, Go, Kaye, and Spasojevic}{Ames
  et~al\mbox{.}}{2010}]%
        {ames2010making}
\bibfield{author}{\bibinfo{person}{Morgan~G Ames}, \bibinfo{person}{Janet Go},
  \bibinfo{person}{Joseph'Jofish' Kaye}, {and} \bibinfo{person}{Mirjana
  Spasojevic}.} \bibinfo{year}{2010}\natexlab{}.
\newblock \showarticletitle{Making love in the network closet: the benefits and
  work of family videochat}. In \bibinfo{booktitle}{\emph{Proceedings of the
  2010 ACM conference on Computer supported cooperative work}}.
  \bibinfo{pages}{145--154}.
\newblock


\bibitem[\protect\citeauthoryear{An, Zhou, Liu, Yin, Du, Huang, and Zhao}{An
  et~al\mbox{.}}{2021}]%
        {an2021vibemoji}
\bibfield{author}{\bibinfo{person}{Pengcheng An}, \bibinfo{person}{Ziqi Zhou},
  \bibinfo{person}{Qing Liu}, \bibinfo{person}{Yifei Yin},
  \bibinfo{person}{Linghao Du}, \bibinfo{person}{Da-Yuan Huang}, {and}
  \bibinfo{person}{Jian Zhao}.} \bibinfo{year}{2021}\natexlab{}.
\newblock \showarticletitle{VibEmoji: Exploring User-authoring Multi-modal
  Emoticons in Social Communication}.
\newblock \bibinfo{journal}{\emph{arXiv preprint arXiv:2112.13555}}
  (\bibinfo{year}{2021}).
\newblock


\bibitem[\protect\citeauthoryear{{Apple Inc.}}{{Apple Inc.}}{2019}]%
        {scenekit}
\bibfield{author}{\bibinfo{person}{{Apple Inc.}}}
  \bibinfo{year}{2019}\natexlab{}.
\newblock \bibinfo{title}{SceneKit | Apple Developer Documentation}.
\newblock
\newblock
\newblock
\shownote{\url{https://developer.apple.com/documentation/scenekit/}.}


\bibitem[\protect\citeauthoryear{{Apple Inc.}}{{Apple Inc.}}{2020a}]%
        {arkit}
\bibfield{author}{\bibinfo{person}{{Apple Inc.}}}
  \bibinfo{year}{2020}\natexlab{a}.
\newblock \bibinfo{title}{ARKit - Augmented Reality - Apple Developer}.
\newblock
\newblock
\newblock
\shownote{\url{https://developer.apple.com/augmented-reality/arkit/}.}


\bibitem[\protect\citeauthoryear{{Apple Inc.}}{{Apple Inc.}}{2020b}]%
        {spritekit}
\bibfield{author}{\bibinfo{person}{{Apple Inc.}}}
  \bibinfo{year}{2020}\natexlab{b}.
\newblock \bibinfo{title}{SpriteKit - Apple Developer}.
\newblock
\newblock
\newblock
\shownote{\url{https://developer.apple.com/spritekit/}.}


\bibitem[\protect\citeauthoryear{Baishya and Neustaedter}{Baishya and
  Neustaedter}{2017}]%
        {baishya2017your}
\bibfield{author}{\bibinfo{person}{Uddipana Baishya} {and}
  \bibinfo{person}{Carman Neustaedter}.} \bibinfo{year}{2017}\natexlab{}.
\newblock \showarticletitle{In Your Eyes: Anytime, Anywhere Video and Audio
  Streaming for Couples}. In \bibinfo{booktitle}{\emph{Proceedings of the 2017
  ACM Conference on Computer Supported Cooperative Work and Social Computing}}.
  \bibinfo{pages}{84--97}.
\newblock


\bibitem[\protect\citeauthoryear{Billinghurst, Nassani, and
  Reichherzer}{Billinghurst et~al\mbox{.}}{2014}]%
        {billinghurst2014social}
\bibfield{author}{\bibinfo{person}{Mark Billinghurst},
  \bibinfo{person}{Alaeddin Nassani}, {and} \bibinfo{person}{Carolin
  Reichherzer}.} \bibinfo{year}{2014}\natexlab{}.
\newblock \showarticletitle{Social panoramas: using wearable computers to share
  experiences}.
\newblock In \bibinfo{booktitle}{\emph{SIGGRAPH Asia 2014 Mobile Graphics and
  Interactive Applications}}. \bibinfo{pages}{1--1}.
\newblock


\bibitem[\protect\citeauthoryear{Bipat, Bos, Vaish, and
  Monroy-Hern{\'a}ndez}{Bipat et~al\mbox{.}}{2019}]%
        {bipat2019analyzing}
\bibfield{author}{\bibinfo{person}{Taryn Bipat},
  \bibinfo{person}{Maarten~Willem Bos}, \bibinfo{person}{Rajan Vaish}, {and}
  \bibinfo{person}{Andr{\'e}s Monroy-Hern{\'a}ndez}.}
  \bibinfo{year}{2019}\natexlab{}.
\newblock \showarticletitle{Analyzing the use of camera glasses in the wild}.
  In \bibinfo{booktitle}{\emph{Proceedings of the 2019 CHI Conference on Human
  Factors in Computing Systems}}. \bibinfo{pages}{1--8}.
\newblock


\bibitem[\protect\citeauthoryear{Blade}{Blade}{2020}]%
        {VuzixBlade}
Blade \bibinfo{year}{2020}\natexlab{}.
\newblock \bibinfo{title}{{Vuzix Blade}}.
\newblock
\newblock
\newblock
\shownote{\url{https://www.vuzix.com/products/blade-smart-glasses}.}


\bibitem[\protect\citeauthoryear{Carman}{Carman}{2019}]%
        {snapagelens}
\bibfield{author}{\bibinfo{person}{Ashley Carman}.}
  \bibinfo{year}{2019}\natexlab{}.
\newblock \bibinfo{title}{Snapchat launches another aging AR lens to lure
  people back to the app}.
\newblock
\newblock
\newblock
\shownote{\url{https://www.theverge.com/2019/11/21/20974367/snapchat-time-machine-baby-aging-lens-ar-launch}.}


\bibitem[\protect\citeauthoryear{Cramer, Rost, and Holmquist}{Cramer
  et~al\mbox{.}}{2011}]%
        {cramer2011performing}
\bibfield{author}{\bibinfo{person}{Henriette Cramer}, \bibinfo{person}{Mattias
  Rost}, {and} \bibinfo{person}{Lars~Erik Holmquist}.}
  \bibinfo{year}{2011}\natexlab{}.
\newblock \showarticletitle{Performing a check-in: emerging practices, norms
  and'conflicts' in location-sharing using foursquare}. In
  \bibinfo{booktitle}{\emph{Proceedings of the 13th international conference on
  human computer interaction with mobile devices and services}}.
  \bibinfo{pages}{57--66}.
\newblock


\bibitem[\protect\citeauthoryear{Cutrell, Czerwinski, and Horvitz}{Cutrell
  et~al\mbox{.}}{2000}]%
        {cutrell2000effects}
\bibfield{author}{\bibinfo{person}{Edward~B Cutrell}, \bibinfo{person}{Mary
  Czerwinski}, {and} \bibinfo{person}{Eric Horvitz}.}
  \bibinfo{year}{2000}\natexlab{}.
\newblock \showarticletitle{Effects of instant messaging interruptions on
  computing tasks}. In \bibinfo{booktitle}{\emph{CHI'00 extended abstracts on
  Human factors in computing systems}}. \bibinfo{pages}{99--100}.
\newblock


\bibitem[\protect\citeauthoryear{Danieau, Guillo, and Dor{\'e}}{Danieau
  et~al\mbox{.}}{2017}]%
        {danieau2017attention}
\bibfield{author}{\bibinfo{person}{Fabien Danieau}, \bibinfo{person}{Antoine
  Guillo}, {and} \bibinfo{person}{Renaud Dor{\'e}}.}
  \bibinfo{year}{2017}\natexlab{}.
\newblock \showarticletitle{Attention guidance for immersive video content in
  head-mounted displays}. In \bibinfo{booktitle}{\emph{2017 IEEE Virtual
  Reality (VR)}}. IEEE, \bibinfo{pages}{205--206}.
\newblock


\bibitem[\protect\citeauthoryear{Denning, Dehlawi, and Kohno}{Denning
  et~al\mbox{.}}{2014}]%
        {denning2014situ}
\bibfield{author}{\bibinfo{person}{Tamara Denning}, \bibinfo{person}{Zakariya
  Dehlawi}, {and} \bibinfo{person}{Tadayoshi Kohno}.}
  \bibinfo{year}{2014}\natexlab{}.
\newblock \showarticletitle{In situ with bystanders of augmented reality
  glasses: Perspectives on recording and privacy-mediating technologies}. In
  \bibinfo{booktitle}{\emph{Proceedings of the SIGCHI Conference on Human
  Factors in Computing Systems}}. \bibinfo{pages}{2377--2386}.
\newblock


\bibitem[\protect\citeauthoryear{Epley and Schroeder}{Epley and
  Schroeder}{2014}]%
        {epley2014mistakenly}
\bibfield{author}{\bibinfo{person}{Nicholas Epley} {and}
  \bibinfo{person}{Juliana Schroeder}.} \bibinfo{year}{2014}\natexlab{}.
\newblock \showarticletitle{Mistakenly seeking solitude.}
\newblock \bibinfo{journal}{\emph{Journal of Experimental Psychology: General}}
  \bibinfo{volume}{143}, \bibinfo{number}{5} (\bibinfo{year}{2014}),
  \bibinfo{pages}{1980}.
\newblock


\bibitem[\protect\citeauthoryear{{Facebook, Inc.}}{{Facebook, Inc.}}{2020}]%
        {insta}
\bibfield{author}{\bibinfo{person}{{Facebook, Inc.}}}
  \bibinfo{year}{2020}\natexlab{}.
\newblock \bibinfo{title}{Instagram}.
\newblock
\newblock
\newblock
\shownote{\url{https://instagram.com/}.}


\bibitem[\protect\citeauthoryear{FBRooms}{FBRooms}{2020}]%
        {FBRooms}
FBRooms \bibinfo{year}{2020}\natexlab{}.
\newblock \bibinfo{title}{{Facebook Rooms}}.
\newblock
\newblock
\newblock
\shownote{\url{https://about.fb.com/news/2020/04/introducing-messenger-rooms/}.}


\bibitem[\protect\citeauthoryear{Fuchs, Bishop, Arthur, McMillan, Bajcsy, Lee,
  Farid, and Kanade}{Fuchs et~al\mbox{.}}{1994}]%
        {fuchs1994virtual}
\bibfield{author}{\bibinfo{person}{Henry Fuchs}, \bibinfo{person}{Gary Bishop},
  \bibinfo{person}{Kevin Arthur}, \bibinfo{person}{Leonard McMillan},
  \bibinfo{person}{Ruzena Bajcsy}, \bibinfo{person}{Sang Lee},
  \bibinfo{person}{Hany Farid}, {and} \bibinfo{person}{Takeo Kanade}.}
  \bibinfo{year}{1994}\natexlab{}.
\newblock \showarticletitle{Virtual space teleconferencing using a sea of
  cameras}. In \bibinfo{booktitle}{\emph{Proc. First International Conference
  on Medical Robotics and Computer Assisted Surgery}},
  Vol.~\bibinfo{volume}{26}.
\newblock


\bibitem[\protect\citeauthoryear{Gauglitz, Nuernberger, Turk, and
  H{\"o}llerer}{Gauglitz et~al\mbox{.}}{2014a}]%
        {Gauglitz2014}
\bibfield{author}{\bibinfo{person}{Steffen Gauglitz}, \bibinfo{person}{Benjamin
  Nuernberger}, \bibinfo{person}{Matthew Turk}, {and} \bibinfo{person}{Tobias
  H{\"o}llerer}.} \bibinfo{year}{2014}\natexlab{a}.
\newblock \showarticletitle{In Touch with the Remote World: Remote
  Collaboration with Augmented Reality Drawings and Virtual Navigation}. In
  \bibinfo{booktitle}{\emph{Proceedings of the 20th {{ACM Symposium}} on
  {{Virtual Reality Software}} and {{Technology}} - {{VRST}} '14}}.
  \bibinfo{publisher}{{ACM Press}}, \bibinfo{address}{{Edinburgh, Scotland}},
  \bibinfo{pages}{197--205}.
\newblock
\showISBNx{978-1-4503-3253-8}
\urldef\tempurl%
\url{https://doi.org/10.1145/2671015.2671016}
\showDOI{\tempurl}


\bibitem[\protect\citeauthoryear{Gauglitz, Nuernberger, Turk, and
  H{\"o}llerer}{Gauglitz et~al\mbox{.}}{2014b}]%
        {gauglitz2014touch}
\bibfield{author}{\bibinfo{person}{Steffen Gauglitz}, \bibinfo{person}{Benjamin
  Nuernberger}, \bibinfo{person}{Matthew Turk}, {and} \bibinfo{person}{Tobias
  H{\"o}llerer}.} \bibinfo{year}{2014}\natexlab{b}.
\newblock \showarticletitle{In touch with the remote world: Remote
  collaboration with augmented reality drawings and virtual navigation}. In
  \bibinfo{booktitle}{\emph{Proceedings of the 20th ACM Symposium on Virtual
  Reality Software and Technology}}. \bibinfo{pages}{197--205}.
\newblock


\bibitem[\protect\citeauthoryear{Google Glass}{Google Glass}{2019}]%
        {GoogleGlass}
Google Glass \bibinfo{year}{2019}\natexlab{}.
\newblock \bibinfo{title}{{Glass}}.
\newblock
\newblock
\newblock
\shownote{\url{https://developers.google.com/glass/develop/gdk/touch}.}


\bibitem[\protect\citeauthoryear{{Google Inc.}}{{Google Inc.}}{2020}]%
        {firebase}
\bibfield{author}{\bibinfo{person}{{Google Inc.}}}
  \bibinfo{year}{2020}\natexlab{}.
\newblock \bibinfo{title}{Firebase}.
\newblock
\newblock
\newblock
\shownote{\url{https://firebase.google.com/}.}


\bibitem[\protect\citeauthoryear{Graffity}{Graffity}{2020}]%
        {Graffity}
Graffity \bibinfo{year}{2020}\natexlab{}.
\newblock \bibinfo{title}{{Graffity App}}.
\newblock
\newblock
\newblock
\shownote{\url{https://www.youtube.com/watch?v=DVy8SHpTxsA}.}


\bibitem[\protect\citeauthoryear{Greenberg and Neustaedter}{Greenberg and
  Neustaedter}{2013}]%
        {greenberg2013shared}
\bibfield{author}{\bibinfo{person}{Saul Greenberg} {and}
  \bibinfo{person}{Carman Neustaedter}.} \bibinfo{year}{2013}\natexlab{}.
\newblock \showarticletitle{Shared living, experiences, and intimacy over video
  chat in long distance relationships}.
\newblock In \bibinfo{booktitle}{\emph{Connecting families}}.
  \bibinfo{publisher}{Springer}, \bibinfo{pages}{37--53}.
\newblock


\bibitem[\protect\citeauthoryear{Hashimoto, Phitayakorn, Fernandez-del
  Castillo, and Meireles}{Hashimoto et~al\mbox{.}}{2016}]%
        {hashimoto2016blinded}
\bibfield{author}{\bibinfo{person}{Daniel~A Hashimoto}, \bibinfo{person}{Roy
  Phitayakorn}, \bibinfo{person}{Carlos Fernandez-del Castillo}, {and}
  \bibinfo{person}{Ozanan Meireles}.} \bibinfo{year}{2016}\natexlab{}.
\newblock \showarticletitle{A blinded assessment of video quality in wearable
  technology for telementoring in open surgery: the Google Glass experience}.
\newblock \bibinfo{journal}{\emph{Surgical endoscopy}} \bibinfo{volume}{30},
  \bibinfo{number}{1} (\bibinfo{year}{2016}), \bibinfo{pages}{372--378}.
\newblock


\bibitem[\protect\citeauthoryear{He, Du, and Perlin}{He et~al\mbox{.}}{2020}]%
        {he2020collabovr}
\bibfield{author}{\bibinfo{person}{Zhenyi He}, \bibinfo{person}{Ruofei Du},
  {and} \bibinfo{person}{Ken Perlin}.} \bibinfo{year}{2020}\natexlab{}.
\newblock \showarticletitle{CollaboVR: A Reconfigurable Framework for Creative
  Collaboration in Virtual Reality}. In \bibinfo{booktitle}{\emph{2020 IEEE
  International Symposium on Mixed and Augmented Reality (ISMAR)}}. IEEE,
  \bibinfo{pages}{542--554}.
\newblock


\bibitem[\protect\citeauthoryear{Henderson and Feiner}{Henderson and
  Feiner}{2011}]%
        {henderson2011augmented}
\bibfield{author}{\bibinfo{person}{Steven~J Henderson} {and}
  \bibinfo{person}{Steven~K Feiner}.} \bibinfo{year}{2011}\natexlab{}.
\newblock \showarticletitle{Augmented reality in the psychomotor phase of a
  procedural task}. In \bibinfo{booktitle}{\emph{2011 10th IEEE International
  Symposium on Mixed and Augmented Reality}}. IEEE, \bibinfo{pages}{191--200}.
\newblock


\bibitem[\protect\citeauthoryear{H{\"o}llerer, Feiner, Terauchi, Rashid, and
  Hallaway}{H{\"o}llerer et~al\mbox{.}}{1999}]%
        {hollerer1999exploring}
\bibfield{author}{\bibinfo{person}{Tobias H{\"o}llerer},
  \bibinfo{person}{Steven Feiner}, \bibinfo{person}{Tachio Terauchi},
  \bibinfo{person}{Gus Rashid}, {and} \bibinfo{person}{Drexel Hallaway}.}
  \bibinfo{year}{1999}\natexlab{}.
\newblock \showarticletitle{Exploring MARS: developing indoor and outdoor user
  interfaces to a mobile augmented reality system}.
\newblock \bibinfo{journal}{\emph{Computers \& Graphics}} \bibinfo{volume}{23},
  \bibinfo{number}{6} (\bibinfo{year}{1999}), \bibinfo{pages}{779--785}.
\newblock


\bibitem[\protect\citeauthoryear{Houseparty}{Houseparty}{2020}]%
        {houseparty}
Houseparty \bibinfo{year}{2020}\natexlab{}.
\newblock \bibinfo{title}{{Houseparty}}.
\newblock
\newblock
\newblock
\shownote{\url{https://houseparty.com/}.}


\bibitem[\protect\citeauthoryear{Inc.}{Inc.}{2019}]%
        {focalssetup}
\bibfield{author}{\bibinfo{person}{North Inc.}}
  \bibinfo{year}{2019}\natexlab{}.
\newblock \bibinfo{title}{{Focals by North}}.
\newblock
\newblock
\newblock
\shownote{\url{https://support.bynorth.com/}.}


\bibitem[\protect\citeauthoryear{Inc.}{Inc.}{2021}]%
        {Spectacles}
\bibfield{author}{\bibinfo{person}{Snap Inc.}} \bibinfo{year}{2021}\natexlab{}.
\newblock \bibinfo{title}{{Next Generation Spectacles}}.
\newblock
\newblock
\newblock
\shownote{\url{https://www.spectacles.com/ca-en/new-spectacles/}.}


\bibitem[\protect\citeauthoryear{Jo and Hwang}{Jo and Hwang}{2013a}]%
        {Jo2013}
\bibfield{author}{\bibinfo{person}{Hyungeun Jo} {and} \bibinfo{person}{Sungjae
  Hwang}.} \bibinfo{year}{2013}\natexlab{a}.
\newblock \showarticletitle{Chili: Viewpoint Control and on-Video Drawing for
  Mobile Video Calls}. In \bibinfo{booktitle}{\emph{{{CHI}} '13 {{Extended
  Abstracts}} on {{Human Factors}} in {{Computing Systems}} on - {{CHI EA}}
  '13}}. \bibinfo{publisher}{{ACM Press}}, \bibinfo{address}{{Paris, France}},
  \bibinfo{pages}{1425}.
\newblock
\showISBNx{978-1-4503-1952-2}
\urldef\tempurl%
\url{https://doi.org/10.1145/2468356.2468610}
\showDOI{\tempurl}


\bibitem[\protect\citeauthoryear{Jo and Hwang}{Jo and Hwang}{2013b}]%
        {jo2013chili}
\bibfield{author}{\bibinfo{person}{Hyungeun Jo} {and} \bibinfo{person}{Sungjae
  Hwang}.} \bibinfo{year}{2013}\natexlab{b}.
\newblock \showarticletitle{Chili: viewpoint control and on-video drawing for
  mobile video calls}.
\newblock In \bibinfo{booktitle}{\emph{CHI'13 Extended Abstracts on Human
  Factors in Computing Systems}}. \bibinfo{pages}{1425--1430}.
\newblock


\bibitem[\protect\citeauthoryear{Jones and O'Neill}{Jones and O'Neill}{2011}]%
        {jones2011contextual}
\bibfield{author}{\bibinfo{person}{Simon Jones} {and} \bibinfo{person}{Eamonn
  O'Neill}.} \bibinfo{year}{2011}\natexlab{}.
\newblock \showarticletitle{Contextual dynamics of group-based sharing
  decisions}. In \bibinfo{booktitle}{\emph{Proceedings of the SIGCHI Conference
  on Human Factors in Computing Systems}}. \bibinfo{pages}{1777--1786}.
\newblock


\bibitem[\protect\citeauthoryear{Judge and Neustaedter}{Judge and
  Neustaedter}{2010}]%
        {judge2010sharing}
\bibfield{author}{\bibinfo{person}{Tejinder~K Judge} {and}
  \bibinfo{person}{Carman Neustaedter}.} \bibinfo{year}{2010}\natexlab{}.
\newblock \showarticletitle{Sharing conversation and sharing life: video
  conferencing in the home}. In \bibinfo{booktitle}{\emph{Proceedings of the
  SIGCHI Conference on Human Factors in Computing Systems}}.
  \bibinfo{pages}{655--658}.
\newblock


\bibitem[\protect\citeauthoryear{Kaye}{Kaye}{2006}]%
        {kaye2006just}
\bibfield{author}{\bibinfo{person}{Joseph'Jofish' Kaye}.}
  \bibinfo{year}{2006}\natexlab{}.
\newblock \showarticletitle{I just clicked to say I love you: rich evaluations
  of minimal communication}. In \bibinfo{booktitle}{\emph{CHI'06 extended
  abstracts on human factors in computing systems}}. \bibinfo{pages}{363--368}.
\newblock


\bibitem[\protect\citeauthoryear{Kelly and Watts}{Kelly and Watts}{2015}]%
        {kelly2015characterising}
\bibfield{author}{\bibinfo{person}{Ryan Kelly} {and} \bibinfo{person}{Leon
  Watts}.} \bibinfo{year}{2015}\natexlab{}.
\newblock \showarticletitle{Characterising the inventive appropriation of emoji
  as relationally meaningful in mediated close personal relationships}.
\newblock \bibinfo{journal}{\emph{Experiences of technology appropriation:
  Unanticipated users, usage, circumstances, and design}}  \bibinfo{volume}{2}
  (\bibinfo{year}{2015}).
\newblock


\bibitem[\protect\citeauthoryear{Khandekar, Higg, Bian, Won~Ryu, O.~Talton~Iii,
  and Kumar}{Khandekar et~al\mbox{.}}{2019}]%
        {khandekar2019opico}
\bibfield{author}{\bibinfo{person}{Sujay Khandekar}, \bibinfo{person}{Joseph
  Higg}, \bibinfo{person}{Yuanzhe Bian}, \bibinfo{person}{Chae Won~Ryu},
  \bibinfo{person}{Jerry O.~Talton~Iii}, {and} \bibinfo{person}{Ranjitha
  Kumar}.} \bibinfo{year}{2019}\natexlab{}.
\newblock \showarticletitle{Opico: a study of emoji-first communication in a
  mobile social app}. In \bibinfo{booktitle}{\emph{Companion Proceedings of The
  2019 World Wide Web Conference}}. \bibinfo{pages}{450--458}.
\newblock


\bibitem[\protect\citeauthoryear{Kim, Bolton, Girouard, Cooperstock, and
  Vertegaal}{Kim et~al\mbox{.}}{2012}]%
        {kim2012telehuman}
\bibfield{author}{\bibinfo{person}{Kibum Kim}, \bibinfo{person}{John Bolton},
  \bibinfo{person}{Audrey Girouard}, \bibinfo{person}{Jeremy Cooperstock},
  {and} \bibinfo{person}{Roel Vertegaal}.} \bibinfo{year}{2012}\natexlab{}.
\newblock \showarticletitle{Telehuman: effects of 3d perspective on gaze and
  pose estimation with a life-size cylindrical telepresence pod}. In
  \bibinfo{booktitle}{\emph{Proceedings of the SIGCHI Conference on Human
  Factors in Computing Systems}}. \bibinfo{pages}{2531--2540}.
\newblock


\bibitem[\protect\citeauthoryear{Kim, Lee, Huang, Kim, Woo, and
  Billinghurst}{Kim et~al\mbox{.}}{2019a}]%
        {Kim2019}
\bibfield{author}{\bibinfo{person}{Seungwon Kim}, \bibinfo{person}{Gun Lee},
  \bibinfo{person}{Weidong Huang}, \bibinfo{person}{Hayun Kim},
  \bibinfo{person}{Woontack Woo}, {and} \bibinfo{person}{Mark Billinghurst}.}
  \bibinfo{year}{2019}\natexlab{a}.
\newblock \showarticletitle{Evaluating the {{Combination}} of {{Visual
  Communication Cues}} for {{HMD}}-Based {{Mixed Reality Remote
  Collaboration}}}. In \bibinfo{booktitle}{\emph{Proceedings of the 2019 {{CHI
  Conference}} on {{Human Factors}} in {{Computing Systems}} - {{CHI}} '19}}.
  \bibinfo{publisher}{{ACM Press}}, \bibinfo{address}{{Glasgow, Scotland Uk}},
  \bibinfo{pages}{1--13}.
\newblock
\showISBNx{978-1-4503-5970-2}
\urldef\tempurl%
\url{https://doi.org/10.1145/3290605.3300403}
\showDOI{\tempurl}


\bibitem[\protect\citeauthoryear{Kim, Lee, Huang, Kim, Woo, and
  Billinghurst}{Kim et~al\mbox{.}}{2019b}]%
        {kim2019evaluating}
\bibfield{author}{\bibinfo{person}{Seungwon Kim}, \bibinfo{person}{Gun Lee},
  \bibinfo{person}{Weidong Huang}, \bibinfo{person}{Hayun Kim},
  \bibinfo{person}{Woontack Woo}, {and} \bibinfo{person}{Mark Billinghurst}.}
  \bibinfo{year}{2019}\natexlab{b}.
\newblock \showarticletitle{Evaluating the combination of visual communication
  cues for HMD-based mixed reality remote collaboration}. In
  \bibinfo{booktitle}{\emph{Proceedings of the 2019 CHI conference on human
  factors in computing systems}}. \bibinfo{pages}{1--13}.
\newblock


\bibitem[\protect\citeauthoryear{Krawczyk, Myszkowski, and Seidel}{Krawczyk
  et~al\mbox{.}}{2005}]%
        {krawczyk2005perceptual}
\bibfield{author}{\bibinfo{person}{Grzegorz Krawczyk}, \bibinfo{person}{Karol
  Myszkowski}, {and} \bibinfo{person}{Hans-Peter Seidel}.}
  \bibinfo{year}{2005}\natexlab{}.
\newblock \showarticletitle{Perceptual effects in real-time tone mapping}. In
  \bibinfo{booktitle}{\emph{Proceedings of the 21st spring conference on
  Computer graphics}}. \bibinfo{pages}{195--202}.
\newblock


\bibitem[\protect\citeauthoryear{Kun, Meulen, and Janssen}{Kun
  et~al\mbox{.}}{2019}]%
        {kun2019calling}
\bibfield{author}{\bibinfo{person}{Andrew~L Kun}, \bibinfo{person}{Hidde
  van~der Meulen}, {and} \bibinfo{person}{Christian~P Janssen}.}
  \bibinfo{year}{2019}\natexlab{}.
\newblock \showarticletitle{Calling while driving using augmented reality:
  Blessing or curse?}
\newblock \bibinfo{journal}{\emph{PRESENCE: Virtual and Augmented Reality}}
  \bibinfo{volume}{27}, \bibinfo{number}{1} (\bibinfo{year}{2019}),
  \bibinfo{pages}{1--14}.
\newblock


\bibitem[\protect\citeauthoryear{Kuzuoka, Kosuge, and Tanaka}{Kuzuoka
  et~al\mbox{.}}{1994}]%
        {kuzuoka1994gesturecam}
\bibfield{author}{\bibinfo{person}{Hideaki Kuzuoka}, \bibinfo{person}{Toshio
  Kosuge}, {and} \bibinfo{person}{Masatomo Tanaka}.}
  \bibinfo{year}{1994}\natexlab{}.
\newblock \showarticletitle{GestureCam: A video communication system for
  sympathetic remote collaboration}. In \bibinfo{booktitle}{\emph{Proceedings
  of the 1994 ACM conference on Computer supported cooperative work}}.
  \bibinfo{pages}{35--43}.
\newblock


\bibitem[\protect\citeauthoryear{Kyt{\"o}, Hirskyj-Douglas, and
  McGookin}{Kyt{\"o} et~al\mbox{.}}{2021}]%
        {kyto2021strangers}
\bibfield{author}{\bibinfo{person}{Mikko Kyt{\"o}}, \bibinfo{person}{Ilyena
  Hirskyj-Douglas}, {and} \bibinfo{person}{David McGookin}.}
  \bibinfo{year}{2021}\natexlab{}.
\newblock \showarticletitle{From Strangers to Friends: Augmenting Face-to-face
  Interactions with Faceted Digital Self-Presentations}. In
  \bibinfo{booktitle}{\emph{Augmented Humans Conference 2021}}.
  \bibinfo{pages}{192--203}.
\newblock


\bibitem[\protect\citeauthoryear{Lederer, Dey, and Mankoff}{Lederer
  et~al\mbox{.}}{2002}]%
        {lederer2002conceptual}
\bibfield{author}{\bibinfo{person}{Scott Lederer}, \bibinfo{person}{Anind~K
  Dey}, {and} \bibinfo{person}{Jennifer Mankoff}.}
  \bibinfo{year}{2002}\natexlab{}.
\newblock \bibinfo{booktitle}{\emph{A conceptual model and a metaphor of
  everyday privacy in ubiquitous computing environments}}.
\newblock \bibinfo{publisher}{Computer Science Division, University of
  California}.
\newblock


\bibitem[\protect\citeauthoryear{Louie, Garg, Werner, Sun, Gergle, and
  Zhang}{Louie et~al\mbox{.}}{2021}]%
        {louie2021opportunistic}
\bibfield{author}{\bibinfo{person}{Ryan Louie}, \bibinfo{person}{Kapil Garg},
  \bibinfo{person}{Jennie Werner}, \bibinfo{person}{Allison Sun},
  \bibinfo{person}{Darren Gergle}, {and} \bibinfo{person}{Haoqi Zhang}.}
  \bibinfo{year}{2021}\natexlab{}.
\newblock \showarticletitle{Opportunistic Collective Experiences: Identifying
  Shared Situations and Structuring Shared Activities at Distance}.
\newblock \bibinfo{journal}{\emph{Proceedings of the ACM on Human-Computer
  Interaction}} \bibinfo{volume}{4}, \bibinfo{number}{CSCW3}
  (\bibinfo{year}{2021}), \bibinfo{pages}{1--32}.
\newblock


\bibitem[\protect\citeauthoryear{Matsuhashi, Kanamoto, and Kurokawa}{Matsuhashi
  et~al\mbox{.}}{2020}]%
        {matsuhashi2020thermal}
\bibfield{author}{\bibinfo{person}{Kodai Matsuhashi}, \bibinfo{person}{Toshiki
  Kanamoto}, {and} \bibinfo{person}{Atsushi Kurokawa}.}
  \bibinfo{year}{2020}\natexlab{}.
\newblock \showarticletitle{Thermal model and countermeasures for future smart
  glasses}.
\newblock \bibinfo{journal}{\emph{Sensors}} \bibinfo{volume}{20},
  \bibinfo{number}{5} (\bibinfo{year}{2020}), \bibinfo{pages}{1446}.
\newblock


\bibitem[\protect\citeauthoryear{M{\"u}ller, Langlotz, and
  Regenbrecht}{M{\"u}ller et~al\mbox{.}}{2016}]%
        {muller2016panovc}
\bibfield{author}{\bibinfo{person}{J{\"o}rg M{\"u}ller},
  \bibinfo{person}{Tobias Langlotz}, {and} \bibinfo{person}{Holger
  Regenbrecht}.} \bibinfo{year}{2016}\natexlab{}.
\newblock \showarticletitle{PanoVC: Pervasive telepresence using mobile
  phones}. In \bibinfo{booktitle}{\emph{2016 IEEE International Conference on
  Pervasive Computing and Communications (PerCom)}}. IEEE,
  \bibinfo{pages}{1--10}.
\newblock


\bibitem[\protect\citeauthoryear{Nardi, Whittaker, and Bradner}{Nardi
  et~al\mbox{.}}{2000}]%
        {nardi2000interaction}
\bibfield{author}{\bibinfo{person}{Bonnie~A Nardi}, \bibinfo{person}{Steve
  Whittaker}, {and} \bibinfo{person}{Erin Bradner}.}
  \bibinfo{year}{2000}\natexlab{}.
\newblock \showarticletitle{Interaction and outeraction: instant messaging in
  action}. In \bibinfo{booktitle}{\emph{Proceedings of the 2000 ACM conference
  on Computer supported cooperative work}}. \bibinfo{pages}{79--88}.
\newblock


\bibitem[\protect\citeauthoryear{Nassani, Kim, Lee, Billinghurst, Langlotz, and
  Lindeman}{Nassani et~al\mbox{.}}{2016a}]%
        {Nassani2016}
\bibfield{author}{\bibinfo{person}{Alaeddin Nassani}, \bibinfo{person}{Hyungon
  Kim}, \bibinfo{person}{Gun Lee}, \bibinfo{person}{Mark Billinghurst},
  \bibinfo{person}{Tobias Langlotz}, {and} \bibinfo{person}{Robert~W.
  Lindeman}.} \bibinfo{year}{2016}\natexlab{a}.
\newblock \showarticletitle{Augmented Reality Annotation for Social Video
  Sharing}. In \bibinfo{booktitle}{\emph{{{SIGGRAPH ASIA}} 2016 {{Mobile
  Graphics}} and {{Interactive Applications}} on - {{SA}} '16}}.
  \bibinfo{publisher}{{ACM Press}}, \bibinfo{address}{{Macau}},
  \bibinfo{pages}{1--5}.
\newblock
\showISBNx{978-1-4503-4551-4}
\urldef\tempurl%
\url{https://doi.org/10.1145/2999508.2999529}
\showDOI{\tempurl}


\bibitem[\protect\citeauthoryear{Nassani, Kim, Lee, Billinghurst, Langlotz, and
  Lindeman}{Nassani et~al\mbox{.}}{2016b}]%
        {nassani2016augmented}
\bibfield{author}{\bibinfo{person}{Alaeddin Nassani}, \bibinfo{person}{Hyungon
  Kim}, \bibinfo{person}{Gun Lee}, \bibinfo{person}{Mark Billinghurst},
  \bibinfo{person}{Tobias Langlotz}, {and} \bibinfo{person}{Robert~W
  Lindeman}.} \bibinfo{year}{2016}\natexlab{b}.
\newblock \showarticletitle{Augmented reality annotation for social video
  sharing}.
\newblock In \bibinfo{booktitle}{\emph{SIGGRAPH ASIA 2016 Mobile Graphics and
  Interactive Applications}}. \bibinfo{pages}{1--5}.
\newblock


\bibitem[\protect\citeauthoryear{Neustaedter, Oduor, Venolia, and
  Judge}{Neustaedter et~al\mbox{.}}{2012}]%
        {neustaedter2012moving}
\bibfield{author}{\bibinfo{person}{Carman Neustaedter}, \bibinfo{person}{Erick
  Oduor}, \bibinfo{person}{Gina Venolia}, {and} \bibinfo{person}{Tejinder~K
  Judge}.} \bibinfo{year}{2012}\natexlab{}.
\newblock \showarticletitle{Moving beyond talking heads to shared experiences:
  the future of personal video communication}. In
  \bibinfo{booktitle}{\emph{Proceedings of the 17th ACM international
  conference on Supporting group work}}. \bibinfo{pages}{327--330}.
\newblock


\bibitem[\protect\citeauthoryear{Neustaedter, Pang, Forghani, Oduor, Hillman,
  Judge, Massimi, and Greenberg}{Neustaedter et~al\mbox{.}}{2015}]%
        {neustaedter2015sharing}
\bibfield{author}{\bibinfo{person}{Carman Neustaedter},
  \bibinfo{person}{Carolyn Pang}, \bibinfo{person}{Azadeh Forghani},
  \bibinfo{person}{Erick Oduor}, \bibinfo{person}{Serena Hillman},
  \bibinfo{person}{Tejinder~K Judge}, \bibinfo{person}{Michael Massimi}, {and}
  \bibinfo{person}{Saul Greenberg}.} \bibinfo{year}{2015}\natexlab{}.
\newblock \showarticletitle{Sharing domestic life through long-term video
  connections}.
\newblock \bibinfo{journal}{\emph{ACM Transactions on Computer-Human
  Interaction (TOCHI)}} \bibinfo{volume}{22}, \bibinfo{number}{1}
  (\bibinfo{year}{2015}), \bibinfo{pages}{1--29}.
\newblock


\bibitem[\protect\citeauthoryear{Neustaedter, Procyk, Chua, Forghani, and
  Pang}{Neustaedter et~al\mbox{.}}{2020}]%
        {neustaedter2020mobile}
\bibfield{author}{\bibinfo{person}{Carman Neustaedter}, \bibinfo{person}{Jason
  Procyk}, \bibinfo{person}{Anezka Chua}, \bibinfo{person}{Azadeh Forghani},
  {and} \bibinfo{person}{Carolyn Pang}.} \bibinfo{year}{2020}\natexlab{}.
\newblock \showarticletitle{Mobile video conferencing for sharing outdoor
  leisure activities over distance}.
\newblock \bibinfo{journal}{\emph{Human--Computer Interaction}}
  \bibinfo{volume}{35}, \bibinfo{number}{2} (\bibinfo{year}{2020}),
  \bibinfo{pages}{103--142}.
\newblock


\bibitem[\protect\citeauthoryear{Newton}{Newton}{2018}]%
        {snapfacefilter}
\bibfield{author}{\bibinfo{person}{Casey Newton}.}
  \bibinfo{year}{2018}\natexlab{}.
\newblock \bibinfo{title}{You can now build your own face filter for Snapchat}.
\newblock
\newblock
\newblock
\shownote{\url{https://www.theverge.com/2018/4/17/17245008/snapchat-face-filter-creation-lens-studio-giphy-discover}.}


\bibitem[\protect\citeauthoryear{Next-Gen Spectacles}{Next-Gen
  Spectacles}{2021}]%
        {SnapSpectacles}
Next-Gen Spectacles \bibinfo{year}{2021}\natexlab{}.
\newblock \bibinfo{title}{{Snap Spectacles}}.
\newblock
\newblock
\newblock
\shownote{\url{https://www.spectacles.com/ca-en/new-spectacles}.}


\bibitem[\protect\citeauthoryear{Nreal}{Nreal}{2020}]%
        {nreal}
Nreal \bibinfo{year}{2020}\natexlab{}.
\newblock \bibinfo{title}{{NREAL AI}}.
\newblock
\newblock
\newblock
\shownote{\url{https://www.nreal.ai/}.}


\bibitem[\protect\citeauthoryear{Oda, Elvezio, Sukan, Feiner, and Tversky}{Oda
  et~al\mbox{.}}{2015}]%
        {oda2015virtual}
\bibfield{author}{\bibinfo{person}{Ohan Oda}, \bibinfo{person}{Carmine
  Elvezio}, \bibinfo{person}{Mengu Sukan}, \bibinfo{person}{Steven Feiner},
  {and} \bibinfo{person}{Barbara Tversky}.} \bibinfo{year}{2015}\natexlab{}.
\newblock \showarticletitle{Virtual replicas for remote assistance in virtual
  and augmented reality}. In \bibinfo{booktitle}{\emph{Proceedings of the 28th
  Annual ACM Symposium on User Interface Software \& Technology}}.
  \bibinfo{pages}{405--415}.
\newblock


\bibitem[\protect\citeauthoryear{O'Hara, Black, and Lipson}{O'Hara
  et~al\mbox{.}}{2006}]%
        {o2006everyday}
\bibfield{author}{\bibinfo{person}{Kenton O'Hara}, \bibinfo{person}{Alison
  Black}, {and} \bibinfo{person}{Matthew Lipson}.}
  \bibinfo{year}{2006}\natexlab{}.
\newblock \showarticletitle{Everyday practices with mobile video telephony}. In
  \bibinfo{booktitle}{\emph{Proceedings of the SIGCHI conference on Human
  Factors in computing systems}}. \bibinfo{pages}{871--880}.
\newblock


\bibitem[\protect\citeauthoryear{Orts-Escolano, Rhemann, Fanello, Chang,
  Kowdle, Degtyarev, Kim, Davidson, Khamis, Dou, et~al\mbox{.}}{Orts-Escolano
  et~al\mbox{.}}{2016}]%
        {orts2016holoportation}
\bibfield{author}{\bibinfo{person}{Sergio Orts-Escolano},
  \bibinfo{person}{Christoph Rhemann}, \bibinfo{person}{Sean Fanello},
  \bibinfo{person}{Wayne Chang}, \bibinfo{person}{Adarsh Kowdle},
  \bibinfo{person}{Yury Degtyarev}, \bibinfo{person}{David Kim},
  \bibinfo{person}{Philip~L Davidson}, \bibinfo{person}{Sameh Khamis},
  \bibinfo{person}{Mingsong Dou}, {et~al\mbox{.}}}
  \bibinfo{year}{2016}\natexlab{}.
\newblock \showarticletitle{Holoportation: Virtual 3d teleportation in
  real-time}. In \bibinfo{booktitle}{\emph{Proceedings of the 29th annual
  symposium on user interface software and technology}}.
  \bibinfo{pages}{741--754}.
\newblock


\bibitem[\protect\citeauthoryear{Oulasvirta, Tamminen, Roto, and
  Kuorelahti}{Oulasvirta et~al\mbox{.}}{2005}]%
        {oulasvirta2005interaction}
\bibfield{author}{\bibinfo{person}{Antti Oulasvirta}, \bibinfo{person}{Sakari
  Tamminen}, \bibinfo{person}{Virpi Roto}, {and} \bibinfo{person}{Jaana
  Kuorelahti}.} \bibinfo{year}{2005}\natexlab{}.
\newblock \showarticletitle{Interaction in 4-second bursts: the fragmented
  nature of attentional resources in mobile HCI}. In
  \bibinfo{booktitle}{\emph{Proceedings of the SIGCHI conference on Human
  factors in computing systems}}. \bibinfo{pages}{919--928}.
\newblock


\bibitem[\protect\citeauthoryear{Pfeil, Chatlani, LaViola~Jr, and
  Wisniewski}{Pfeil et~al\mbox{.}}{2021}]%
        {pfeil2021bridging}
\bibfield{author}{\bibinfo{person}{Kevin~P Pfeil}, \bibinfo{person}{Neeraj
  Chatlani}, \bibinfo{person}{Joseph~J LaViola~Jr}, {and}
  \bibinfo{person}{Pamela Wisniewski}.} \bibinfo{year}{2021}\natexlab{}.
\newblock \showarticletitle{Bridging the Socio-Technical Gaps in Body-worn
  Interpersonal Live-Streaming Telepresence through a Critical Review of the
  Literature}.
\newblock \bibinfo{journal}{\emph{Proceedings of the ACM on Human-Computer
  Interaction}} \bibinfo{volume}{5}, \bibinfo{number}{CSCW1}
  (\bibinfo{year}{2021}), \bibinfo{pages}{1--39}.
\newblock


\bibitem[\protect\citeauthoryear{Pfleging, Schneegass, and Schmidt}{Pfleging
  et~al\mbox{.}}{2013}]%
        {pfleging2013exploring}
\bibfield{author}{\bibinfo{person}{Bastian Pfleging}, \bibinfo{person}{Stefan
  Schneegass}, {and} \bibinfo{person}{Albrecht Schmidt}.}
  \bibinfo{year}{2013}\natexlab{}.
\newblock \showarticletitle{Exploring user expectations for context and road
  video sharing while calling and driving}. In
  \bibinfo{booktitle}{\emph{Proceedings of the 5th International Conference on
  Automotive User Interfaces and Interactive Vehicular Applications}}.
  \bibinfo{pages}{132--139}.
\newblock


\bibitem[\protect\citeauthoryear{Piumsomboon, Lee, Hart, Ens, Lindeman, Thomas,
  and Billinghurst}{Piumsomboon et~al\mbox{.}}{2018}]%
        {piumsomboon2018mini}
\bibfield{author}{\bibinfo{person}{Thammathip Piumsomboon},
  \bibinfo{person}{Gun~A Lee}, \bibinfo{person}{Jonathon~D Hart},
  \bibinfo{person}{Barrett Ens}, \bibinfo{person}{Robert~W Lindeman},
  \bibinfo{person}{Bruce~H Thomas}, {and} \bibinfo{person}{Mark Billinghurst}.}
  \bibinfo{year}{2018}\natexlab{}.
\newblock \showarticletitle{Mini-me: An adaptive avatar for mixed reality
  remote collaboration}. In \bibinfo{booktitle}{\emph{Proceedings of the 2018
  CHI conference on human factors in computing systems}}.
  \bibinfo{pages}{1--13}.
\newblock


\bibitem[\protect\citeauthoryear{Piumsomboon, Lee, Lee, Dey, and
  Billinghurst}{Piumsomboon et~al\mbox{.}}{2017}]%
        {Piumsomboon2017}
\bibfield{author}{\bibinfo{person}{Thammathip Piumsomboon},
  \bibinfo{person}{Youngho Lee}, \bibinfo{person}{Gun~A. Lee},
  \bibinfo{person}{Arindam Dey}, {and} \bibinfo{person}{Mark Billinghurst}.}
  \bibinfo{year}{2017}\natexlab{}.
\newblock \showarticletitle{Empathic {{Mixed Reality}}: {{Sharing What You
  Feel}} and {{Interacting}} with {{What You See}}}. In
  \bibinfo{booktitle}{\emph{2017 {{International Symposium}} on {{Ubiquitous
  Virtual Reality}} ({{ISUVR}})}}. \bibinfo{publisher}{{IEEE}},
  \bibinfo{address}{{Nara, Japan}}, \bibinfo{pages}{38--41}.
\newblock
\showISBNx{978-1-5386-3091-4}
\urldef\tempurl%
\url{https://doi.org/10.1109/ISUVR.2017.20}
\showDOI{\tempurl}


\bibitem[\protect\citeauthoryear{Procyk, Neustaedter, Pang, Tang, and
  Judge}{Procyk et~al\mbox{.}}{2014}]%
        {procyk2014exploring}
\bibfield{author}{\bibinfo{person}{Jason Procyk}, \bibinfo{person}{Carman
  Neustaedter}, \bibinfo{person}{Carolyn Pang}, \bibinfo{person}{Anthony Tang},
  {and} \bibinfo{person}{Tejinder~K Judge}.} \bibinfo{year}{2014}\natexlab{}.
\newblock \showarticletitle{Exploring video streaming in public settings:
  shared geocaching over distance using mobile video chat}. In
  \bibinfo{booktitle}{\emph{Proceedings of the SIGCHI Conference on Human
  Factors in Computing Systems}}. \bibinfo{pages}{2163--2172}.
\newblock


\bibitem[\protect\citeauthoryear{Ran, Slocum, Gorlatova, and Chen}{Ran
  et~al\mbox{.}}{2019}]%
        {ran2019sharear}
\bibfield{author}{\bibinfo{person}{Xukan Ran}, \bibinfo{person}{Carter Slocum},
  \bibinfo{person}{Maria Gorlatova}, {and} \bibinfo{person}{Jiasi Chen}.}
  \bibinfo{year}{2019}\natexlab{}.
\newblock \showarticletitle{ShareAR: Communication-efficient multi-user mobile
  augmented reality}. In \bibinfo{booktitle}{\emph{Proceedings of the 18th ACM
  Workshop on Hot Topics in Networks}}. \bibinfo{pages}{109--116}.
\newblock


\bibitem[\protect\citeauthoryear{Rixen, Hirzle, Colley, Etzel, Rukzio, and
  Gugenheimer}{Rixen et~al\mbox{.}}{2021}]%
        {rixen2021exploring}
\bibfield{author}{\bibinfo{person}{Jan~Ole Rixen}, \bibinfo{person}{Teresa
  Hirzle}, \bibinfo{person}{Mark Colley}, \bibinfo{person}{Yannick Etzel},
  \bibinfo{person}{Enrico Rukzio}, {and} \bibinfo{person}{Jan Gugenheimer}.}
  \bibinfo{year}{2021}\natexlab{}.
\newblock \showarticletitle{Exploring Augmented Visual Alterations in
  Interpersonal Communication}. In \bibinfo{booktitle}{\emph{Proceedings of the
  2021 CHI Conference on Human Factors in Computing Systems}}.
  \bibinfo{pages}{1--11}.
\newblock


\bibitem[\protect\citeauthoryear{Rokid Inc.}{Rokid Inc.}{2019}]%
        {Rokid}
Rokid Inc. \bibinfo{year}{2019}\natexlab{}.
\newblock \bibinfo{title}{{Rokid Glass}}.
\newblock
\newblock
\newblock
\shownote{\url{https://glass.rokid.com/en/}.}


\bibitem[\protect\citeauthoryear{Ryskeldiev, Cohen, and Herder}{Ryskeldiev
  et~al\mbox{.}}{2018}]%
        {ryskeldiev2018streamspace}
\bibfield{author}{\bibinfo{person}{Bektur Ryskeldiev}, \bibinfo{person}{Michael
  Cohen}, {and} \bibinfo{person}{Jens Herder}.}
  \bibinfo{year}{2018}\natexlab{}.
\newblock \showarticletitle{Streamspace: Pervasive mixed reality telepresence
  for remote collaboration on mobile devices}.
\newblock \bibinfo{journal}{\emph{Journal of Information Processing}}
  \bibinfo{volume}{26} (\bibinfo{year}{2018}), \bibinfo{pages}{177--185}.
\newblock


\bibitem[\protect\citeauthoryear{Schildt, Leinfors, and Barkhuus}{Schildt
  et~al\mbox{.}}{2016}]%
        {schildt2016communication}
\bibfield{author}{\bibinfo{person}{Emily Schildt}, \bibinfo{person}{Martin
  Leinfors}, {and} \bibinfo{person}{Louise Barkhuus}.}
  \bibinfo{year}{2016}\natexlab{}.
\newblock \showarticletitle{Communication, coordination and awareness around
  continuous location sharing}. In \bibinfo{booktitle}{\emph{Proceedings of the
  19th International Conference on Supporting Group Work}}.
  \bibinfo{pages}{257--265}.
\newblock


\bibitem[\protect\citeauthoryear{Sinch}{Sinch}{2020}]%
        {sinch}
\bibfield{author}{\bibinfo{person}{Sinch}.} \bibinfo{year}{2020}\natexlab{}.
\newblock \bibinfo{title}{{Video Calling - Get face to face with our simple
  APIs and SDKs}}.
\newblock
\newblock
\newblock
\shownote{\url{https://www.sinch.com/products/apis/calling/video/}.}


\bibitem[\protect\citeauthoryear{Smoodji}{Smoodji}{2018}]%
        {smoodji}
Smoodji \bibinfo{year}{2018}\natexlab{}.
\newblock \bibinfo{title}{Smoodji Augmented Reality Application}.
\newblock
\newblock
\newblock
\shownote{\url{https://www.octosense.com/project/smoodji-augmented-reality-messaging/}.}


\bibitem[\protect\citeauthoryear{Spectacles}{Spectacles}{2019}]%
        {SnapSpects}
Spectacles \bibinfo{year}{2019}\natexlab{}.
\newblock \bibinfo{title}{{Snap Focals}}.
\newblock
\newblock
\newblock
\shownote{\url{https://www.spectacles.com/ca-en/learn/}.}


\bibitem[\protect\citeauthoryear{Thoravi~Kumaravel, Anderson, Fitzmaurice,
  Hartmann, and Grossman}{Thoravi~Kumaravel et~al\mbox{.}}{2019}]%
        {thoravi2019loki}
\bibfield{author}{\bibinfo{person}{Balasaravanan Thoravi~Kumaravel},
  \bibinfo{person}{Fraser Anderson}, \bibinfo{person}{George Fitzmaurice},
  \bibinfo{person}{Bjoern Hartmann}, {and} \bibinfo{person}{Tovi Grossman}.}
  \bibinfo{year}{2019}\natexlab{}.
\newblock \showarticletitle{Loki: Facilitating remote instruction of physical
  tasks using bi-directional mixed-reality telepresence}. In
  \bibinfo{booktitle}{\emph{Proceedings of the 32nd Annual ACM Symposium on
  User Interface Software and Technology}}. \bibinfo{pages}{161--174}.
\newblock


\bibitem[\protect\citeauthoryear{Xu, Mukawa, Li, Lim, Tan, Chia, Gan, and
  Mandal}{Xu et~al\mbox{.}}{2015}]%
        {xu2015exploring}
\bibfield{author}{\bibinfo{person}{Qianli Xu}, \bibinfo{person}{Michal Mukawa},
  \bibinfo{person}{Liyuan Li}, \bibinfo{person}{Joo~Hwee Lim},
  \bibinfo{person}{Cheston Tan}, \bibinfo{person}{Shue~Ching Chia},
  \bibinfo{person}{Tian Gan}, {and} \bibinfo{person}{Bappaditya Mandal}.}
  \bibinfo{year}{2015}\natexlab{}.
\newblock \showarticletitle{Exploring users' attitudes towards social
  interaction assistance on Google Glass}. In
  \bibinfo{booktitle}{\emph{Proceedings of the 6th Augmented Human
  international conference}}. \bibinfo{pages}{9--12}.
\newblock


\end{thebibliography}

\end{document}